# Toward 100% Spin-Orbit Torque Efficiency with High Spin-Orbital Hall Conductivity Pt-Cr Alloys

*Chen-Yu Hu†, Yu-Fang Chiu†, Chia-Chin Tsai, Chao-Chung Huang, Kuan-Hao Chen, Cheng-Wei Peng, Chien-Min Lee, Ming-Yuan Song, Yen-Lin Huang, Shy-Jay Lin, and Chi-Feng Pai\**

Chen-Yu Hu, Yu-Fang Chiu, Chia-Chin Tsai, Chao-Chung Huang, Kuan-Hao Chen, Prof. Chi-Feng Pai

Department of Materials Science and Engineering, National Taiwan University

Taipei 10617, Taiwan

E-mail: cfpai@ntu.edu.tw

Chien-Min Lee, Ming-Yuan Song, Yen-Lin Huang, Shy-Jay Lin

Corporate Research, Taiwan Semiconductor Manufacturing Company

Hsinchu 30078, Taiwan

Keywords: spin-orbit torques, spin Hall effects, orbital Hall effects, spintronics, SOT-MRAM, magnetic materials

**Abstract**

5$d$ transition metal Pt is the canonical spin Hall material for efficient generation of spin-orbit torques (SOTs) in Pt/ferromagnetic layer (FM) heterostructures. However, for a long while with tremendous engineering endeavors, the damping-like SOT efficiencies ($\xi_{DL}$) of Pt and Pt alloys have still been limited to $\xi_{DL}$ < 0.5. Here we present that with proper alloying elements, particularly 3$d$ transition metals V and Cr, a high spin-orbital Hall conductivity ($\sigma_{SH} \sim 6.5 \times 10^5 \left(\frac{\hbar}{2e}\right) \Omega^{-1} \cdot m^{-1}$) can be developed. Especially for the Cr-doped case, an extremely high $\xi_{DL} \sim$ 0.9 in a $Pt_{0.69}Cr_{0.31}$/Co device can be achieved with a moderate $Pt_{0.69}Cr_{0.31}$ resistivity of $\rho_{xx} \sim 133\ \mu\Omega \cdot cm$. A low critical SOT-driven switching current density of $J_c \sim 3.2 \times 10^6\ A \cdot cm^{-2}$ is also demonstrated. The damping constant ($\alpha$) of the $Pt_{0.69}Cr_{0.31}$/FM structure is also found to be reduced to 0.052 from the pure Pt/FM case of 0.078. The overall high $\sigma_{SH}$, giant $\xi_{DL}$, moderate $\rho_{xx}$, and reduced $\alpha$ of such a Pt-Cr/FM heterostructure makes it promising for versatile extremely low power consumption SOT memory applications.

1. Introduction

The discovery of the giant spin Hall effect (SHE)[1, 2] in 5$d$ transition metal Pt[3, 4] can be considered as one of the milestones in contemporary spintronics research. The sizable SHE in Pt, which can be used to efficiently convert charge current to spin current, has been intensively studied through experiments of various forms: nonlocal spin injection[5], spin-pumping[6], and spin-torque ferromagnetic resonance[3]. Most importantly, magnetization switching[7-9] and oscillations[10] driven by the SHE-induced spin transfer torque, later coined as the spin-orbit torques (SOTs), in Pt/ferromagnetic layer (FM) heterostructures have demonstrated the efficacy of such a magnetization manipulation mechanism for the potential magnetic memory and oscillator applications. To reduce the power consumption for controlling these Pt-based SOT devices and make them more advantageous than their conventional spin-transfer torque (STT) counterparts, it is of great importance to enhance the damping-like SOT (DL-SOT) efficiency ($\xi_{DL}$) of pure-Pt/FM structures (typically ranges from 0.10 to 0.20[3, 11]). This is commonly done by incorporating the Pt layer with dopants such as Au[12] ($\xi_{DL} \sim 0.35$), Pd[13] ($\xi_{DL} \sim 0.26$), MgO[14] ($\xi_{DL} \sim 0.28$), Al[15] ($\xi_{DL} \sim 0.14$), Hf[15] ($\xi_{DL} \sim 0.16$), and Cu[16] ($\xi_{DL} \sim 0.44$). Also, modification on the Pt/FM interface is another method to improve the DL-SOT efficiency, such as introducing a thin NiO insertion layer[17] ($\xi_{DL} \sim 0.8$) or SiO$_2$ capping layer[18] ($\xi_{DL} \sim 2.83$). However, compared to the STT case, where the charge-to-spin conversion is governed by the spin polarization factor ($P$) of the ferromagnetic layer in magnetic tunnel junctions (or spin valves), most of these enhanced $\xi_{DL}$ generally are still smaller than some common values of $P$ (for example, $P_{CoFeB} \sim 0.5^{19}$ to $0.65^{20}$).

In contrast to the conventional SHE mechanism, the orbital Hall effect (OHE) in light $3d$ transition metals, such as Cr, V and Mn, have been theoretically predicted to give rise to sizable orbital Hall conductivities that are comparable to the spin Hall conductivities of the conventional $5d$ transition metals[21-24]. The sizable orbital current generation by $3d$ transition metals has been experimentally verified[25], especially in a Cr/Pt/CoFeB system[26]. The orbital-to-spin angular momentum conversion is considered as an additional pathway to generate spin current and therefore further enhance the SOT efficiency. Here we show that Pt-V and Pt-Cr with optimized $5d$-to-$3d$ transition metal compositions are promising spin Hall alloys that can significantly enhance SOT efficiencies in magnetic heterostructures, in conjunction with FM layers having either perpendicular magnetic anisotropy (PMA) or in-plane magnetic anisotropy (IMA). The SOT efficiencies are systematically characterized by anomalous Hall effect (AHE) loop shift measurements and planar Hall effect (PHE) curve shift measurements for samples with PMA and IMA, respectively. The estimated DL-SOT efficiency is maximized in the Pt$_{0.69}$Cr$_{0.31}$ sample, where $\xi_{DL} \sim 0.9$ (0.89 for the PMA case and 0.94 for the IMA case). The average spin Hall conductivities of these Pt alloys ($\sigma_{SH}^{Pt-Cr}$) are further estimated to be $6.45 \times 10^5 \left(\frac{\hbar}{2e}\right) \Omega^{-1} \cdot m^{-1}$. Spin-torque ferromagnetic resonance (ST-FMR) is also performed to obtain the variation of Gilbert damping constant (α) versus dopant composition, which is reduced from 0.078 (pure Pt) to 0.052 (Pt$_{0.69}$Cr$_{0.31}$). Current-induced SOT switching is demonstrated with a low critical switching current density of $J_c \sim 3.2 \times 10^6$ A·cm$^{-2}$. From these results we conclude that Pt$_{0.69}$Cr$_{0.31}$ is by far the most efficient transition metal alloy in terms of charge-to-spin conversion efficiency, with the lowest projected power consumption per SOT switching.

## 2. Sample preparation and materials characterization

Multilayer structures Pt$_x$Cr$_{1-x}$(5)/Co(1.6)/MgO(2) (PMA), Pt$_x$V$_{1-x}$(5)/Co(1.6)/MgO(2) (PMA), and Pt$_x$Cr$_{1-x}$(3)/CoFeB(2.3)/MgO(2) (IMA) are first deposited onto thermally-oxidized Si substrates by high vacuum magnetron sputtering. The base and working pressures are $1 \times 10^{-8}$ Torr and $1 \times 10^{-3}$ Torr, respectively. The Pt-Cr(V) alloys are prepared by cosputtering from pure Pt and Cr(V) targets (details can be found in the Supporting Information). x represents the atomic percentages of Pt. Numbers in the parentheses represent the nominal thicknesses in nanometers. These films are then patterned into micron-sized Hall bars (dimensions of 5 μm by 60 μm, as depicted in **Figure 1**a) by photolithography with a lift-off process. The saturation magnetization ($M_s$) of the FM layers are determined by vibrating sample magnetometer (VSM) and superconducting quantum interference device magnetometer (SQUID), giving $M_s = 1116$ emu/cm$^3$ with $t_{dead} = 0.49$ nm for the Co case and $M_s = 860$ emu/$cm^3$ with a negligible $t_{dead}$ for the CoFeB case. The resistivities of the Pt$_x$Cr$_{1-x}$ and Pt$_x$V$_{1-x}$ films are further characterized by four-point measurements on the Hall bar devices, giving $\rho_{Pt} = 49.7 \pm 0.01$ μΩ·cm for the pure-Pt-based control samples. As shown in **Figure 1**b, the alloy resistivities increase monotonically with increasing ratio of Cr and V dopants. The resistivities of Co and CoFeB layers are determined separately, giving $\rho_{Co} = 22.6 \pm 1.8$ μΩ·cm and $\rho_{CoFeB} = 134.0 \pm 9.8$ μΩ·cm. Anomalous Hall effect (AHE) measurements on the Hall bar devices are performed to characterize magnetic anisotropy, using the ratio of remnant magnetization ($M_r$) to saturation magnetization ($M_s$) as the figure of merit. As shown in **Figure 1**c, PMA can be preserved in a larger alloying range for the Pt-Cr series than the Pt-V series. **Figure 1**d shows that the out-of-plane coercivity ($H_c$) is slightly enhanced by alloying a small content of Cr and V, which is possibly due to Cr (or V)/Co interface can improve the magnetic anisotropy under optimal condition[27, 28]. As more Cr is incorporated into the Pt-Cr alloy, $H_c$ then starts to decrease due to the reduced interfacial SOC at Pt/Co interface[29].

3. **Spin-orbit torque measurements**

a. DL-SOT efficiency characterization (PMA)

The DL-SOT efficiency of heterostructures with PMA is characterized via anomalous Hall resistance hysteresis loop shift measurements[30], which are performed by applying a dc in-plane current and an external in-plane magnetic field simultaneously, both are along *x*-direction, while measuring out-of-plane anomalous Hall resistance hysteresis loop of the Hall bar devices. The experimental setup is as shown in **Figure 2**a. The charge current flowing in the spin current source (SCS) layer would generate a perpendicular spin current via the SHE and then transfer a DL-SOT to the adjacent FM. The DL-SOT experienced by the FM layer with PMA can be represented by a perpendicular effective field ($H_{\text{eff}}^z$) and therefore manifests itself in the shift of the hysteresis loop center. Representative data of current-induced hysteresis loop shifts are shown in **Figure 2**b. By performing linearly fits to the current-dependent effective field data and extracting $\frac{H_{\text{eff}}^z}{I}$, as shown in **Figure 2**c, the DL-SOT efficiency then can be estimated via[30, 31]:

$$\xi_{\text{DL}} = \left(\frac{2}{\pi}\right)\left(\frac{2e}{\hbar}\right)\mu_0 M_s(t_{\text{FM}} - t_{\text{dead}})w_{\text{eff}} t_{\text{SCS}}(1+s)\left(\frac{H_{\text{eff}}^z}{I}\right), \quad (1)$$

where $e$, $\hbar$ and $\mu_0$ respectively are elementary charge, reduced Planck constant and vacuum permittivity; $t_{FM}$ and $t_{SCS}$ are the nominal thickness of ferromagnetic layer and SCS layer; $w_{eff}$ is the effective channel width with a correction factor of 1.58 ($w_{eff} \approx 1.58 \times w_{nominal}$), which accounts for the effect of lateral current shunting at the Hall cross center[32]; $s$ is the shunting parameter, defined as $s \equiv \frac{I_{FM}}{I_{SCS}} = \frac{t_{FM}\rho_{SCS}}{t_{SCS}\rho_{FM}}$. The correction on effective width of the current channel is necessary to address the exact value of current density[33], and the correction factor 1.58 is adopted according to the same device geometry in a previous work[32], which is determined by comparing the results from the electrical and the magneto-optic Kerr effect (MOKE) hysteresis loop-shift measurements. Note that since we use micron-sized devices, the magnetization switching follows a domain wall motion scenario with chiral Néel wall stabilized by the interfacial Dzyaloshinskii-Moriya interaction (iDMI)[34, 35]. The saturated DL-SOT efficiency can be measured only when the external in-plane magnetic field overcomes the DMI effective field $H_{DMI}$ and aligns the domain wall moment, which is shown in **Figure 2**d.

Compared to the control sample (pure Pt), Pt-V and Pt-Cr both show superior charge-to-spin conversion ability, where $\frac{H_{eff}^z}{I}$ = 27.9 ± 0.35 Oe/mA for $Pt_{0.8}V_{0.2}$, 32.5 ± 0.03 Oe/mA for $Pt_{0.69}Cr_{0.31}$, 13.7 ± 1.57 Oe/mA for pure Pt, respectively. The relation between the concentration of Pt and $\frac{H_{eff}^z}{I}$ is shown in **Figure 3**a. The DMI is found to be attenuated with increasing Cr or V doping, as shown in **Figure 3**b, which is tentatively attributed to the weakening of the interfacial spin-orbit coupling by incorporating $3d$ transition metals in Pt.

From the obtained saturated $\frac{H_{eff}^z}{I}$ and layer resistivities, the calculated $\xi_{DL}$ of the control samples (pure Pt) is 0.22 ± 0.04, which is quite similar to previous reports[12, 13]. In **Figure 3**c, the enhanced $\xi_{DL}$ can reach as high as 0.58 ± 0.05 for $Pt_{0.8}V_{0.2}$ and 0.92 ± 0.07 for $Pt_{0.61}Cr_{0.39}$ within the PMA window ($Pt_{0.69}Cr_{0.31}$ with $\xi_{DL}$ of 0.89 ± 0.06 shows the largest $\frac{H_{eff}^z}{I}$ in the Pt-Cr series). This efficiency enhancement from Cr alloying is also similar to that of a different heterostructure, annealed Pt-Cr/Pt/Co/AlOx[36]. With a much larger $\xi_{DL}$ (close to unity) and smaller $\rho_{xx}$, Pt-Cr alloy should be more advantageous in terms of power consumption than the best reported W[37], a common SCS in recent industrial application.

This tremendous improvement of $\xi_{DL}$ is realized by properly tuning of $\rho_{xx}$, since $\xi_{DL} = T_{int}\theta_{SH} = T_{int}\sigma_{SH}\rho_{xx}$, where $\theta_{SH} \equiv \frac{j_s}{j_e} = \sigma_{SH}\rho_{xx}$ is the internal spin Hall angle[2] representing the ratio between the generated spin current density ($j_s$) and charge current density ($j_e$) flowing in the spin current source, and $T_{int}$ is the interfacial spin transparency factor[11]. **Figure 3**d shows an obvious positive correlation between longitudinal $\rho_{xx}$ and $\xi_{DL}$ within a relatively conductive regime, and the spin Hall conductivity (SHC) of pure Pt ($\sigma_{SH}^{Pt}$) is $4.47 \times 10^5 \left(\frac{\hbar}{2e}\right) \Omega^{-1} \cdot m^{-1}$. The magnitude of this apparent SHC is quite consistent with recent consensus that the SHC of Pt ranges from $5.9 \times 10^5$ to $8 \times 10^6 \left(\frac{\hbar}{2e}\right) \Omega^{-1} \cdot m^{-1}$ [14, 38, 39]. And the average spin Hall conductivity (SHC) of alloys ($\sigma_{SH}^{Pt-Cr(V)}$) are $6.45 \times 10^5 \left(\frac{\hbar}{2e}\right) \Omega^{-1} \cdot m^{-1}$ for Pt-Cr and $6.03 \times 10^5 \left(\frac{\hbar}{2e}\right) \Omega^{-1} \cdot m^{-1}$ for Pt-V. Additionally, if we further consider the spin transmission degradation from spin-memory loss (SML)[40], spin backflow (SBF)[41-43], and possible effects from magnetic dead layer at the Pt/Co interface[17], the actual SHC would be even higher. In the relatively resistive regime, $\xi_{DL}$ is fairly independent of further raise of $\rho_{xx}$ and remains at a high value within the PMA window. This feature combined with the increasing $\rho_{xx}$ contributes to the dropping of $\frac{H_{eff}^z}{I}$ as the Cr concentration goes beyond 0.34 in the Pt-Cr series. It's noteworthy that alloying by light $3d$ transition metals (V and Cr) and heavy $3d$ transition metals (Cu)[16] shows similar improvement on $\xi_{DL}$ in the Pt-X/Co/MgO (X = V, Cr, Cu) structures under the premise of raising $\rho_{xx}$ and preserving

PMA. Considering that by introducing Cr and V into Pt would deficit the average valence electron number, coupled with the fact that Cr and V possess negative spin Hall angles[44], therefore, $\rho_{xx}$ (enhancing scattering)[14] possibly plays a more critical role in tuning $\xi_{DL}$ than the band filling effect (adjusting the Fermi level)[45, 46] in conductive Pt alloys.

b. DL-SOT efficiency characterization (IMA)

To further verify our discovery, planar Hall effect (PHE) curve shift measurement is performed to characterize $\xi_{DL}$ from the IMA devices, as illustrated in **Figure 4**a. Considering that the equilibrium magnetization direction is affected by the effective anisotropy field ($H_k^{eff}$), current-induced field(s) ($\Delta H$) and external field ($H_{ext}$), and under the premise of in-plane magnetization with negligible in-plane anisotropy, the expression of angle-dependent transverse resistance ($R_{xy}$) contributed from anomalous Hall effect (AHE), planar Hall effect (PHE) and anomalous Nernst effect (ANE) is expressed as

$$R_{xy} \sim R_{PHE} \sin 2\varphi_H + \left[ R_{AHE} \left( \frac{\Delta H_{DL}}{-H_k^{eff} + H_{ext}} \right) + I\alpha_{ANE} \nabla T \right] \cos \varphi_H + \qquad (2)$$
$$R_{PHE} \left[ \frac{-2(\Delta H_{FL} + \Delta H_{Oe})}{H_{ext}} \right] (2\cos^3 \varphi_H - \cos \varphi_H),$$

where $R_{PHE(AHE)}$ is half of the maximum resistance difference due to PHE or AHE; $\varphi_H$ is the azimuthal angel of applied external field; $\Delta H_{DL(FL\ or\ Oe)}$ is the current-induced damping(field)-like spin-orbit effective field or Oersted field; $H_k^{eff}$ is the effective anisotropy field; $H_{ext}$ is the external applied field; $I$ is the injected dc current amount; $\alpha_{ANE}$ is the ANE coefficient; $\nabla T$ is the perpendicular thermal gradient between the substrate and the deposited layers (along positive $z$-direction). Then considering the symmetry of each term to the applied current direction ($\pm I$), the average/difference of transverse resistance ($\overline{R_{xy}}$ and $\Delta R_{xy}$) under opposite current direction should respectively contain current-dependent/independent terms, which can be expressed as

$$\overline{R_{xy}} \equiv \tfrac{1}{2}[R_{xy}(+I) + R_{xy}(-I)] = R_{\text{PHE}} \sin 2\varphi_H,$$

$$\Delta R_{xy} \equiv \tfrac{1}{2}[R_{xy}(+I) - R_{xy}(-I)] = \left[R_{\text{AHE}}\left(\frac{\Delta H_{\text{DL}}}{-H_k^{\text{eff}} + H_{\text{ext}}}\right) + I\alpha_{\text{ANE}} \nabla T\right]\cos\varphi_H +$$

$$R_{\text{PHE}}\left[\frac{-2(\Delta H_{\text{FL}} + \Delta H_{\text{Oe}})}{H_{\text{ext}}}\right](2\cos^3\varphi_H - \cos\varphi_H) = R^{\text{DL+ANE}}\cos\varphi_H + \tag{3}$$

$$R^{\text{FL+Oe}}(2\cos^3\varphi_H - \cos\varphi_H),$$

which is analogous to the harmonic measurement[47, 48], where $R_{xy}(\pm I)$ is the measured $R_{xy}$ under $\pm I$. A representative angle-dependence of $\overline{R_{xy}}$ is plotted in **Figure 4**b, showing a typical $\sin 2\varphi_H$ trend originated from the conventional PHE. By decomposing $\Delta R_{xy}$ and extracting the weight of $\cos\varphi_H$ component in the angle-dependent $\Delta R_{xy}$ with $I = \pm 3$ mA, $\Delta H_{\text{DL}}$ can be estimated from the linear fitting under varying $H_{\text{ext}}$, as shown in **Figure 4**c-d, giving $\Delta H_{\text{DL}} \sim 18.1$ Oe for pure Pt and $\Delta H_{\text{DL}} \sim 34.0$ Oe for Pt$_{0.69}$Cr$_{0.31}$, where $H_k^{\text{eff}}$ is -8000 Oe based on the results from SQUID measurement and $R_{\text{AHE}} \sim -\frac{\partial \overline{R_{xy}}}{\partial \theta_H}$ is determined by measuring $R_{xy}$ under scanning $H_{\text{ext}}$ along $x$-$z$ plane ($\varphi_H = 0°$) with a fixed field strength. $\xi_{\text{DL}}$ can thus be estimated by

$$\xi_{\text{DL}} = \left(\frac{2e}{\hbar}\right)\mu_0 M_s (t_{\text{FM}} - t_{\text{dead}}) w_{\text{eff}} t_{\text{SCS}} (1+s)\left(\frac{\Delta H_{\text{DL}}}{I}\right). \tag{4}$$

The estimated $\xi_{DL}$ from PHE curve shift measurement is 0.34 ± 0.05 for pure Pt and 0.94 ± 0.14 for Pt$_{0.69}$Cr$_{0.31}$ IMA devices, which are quite consistent with that of the PMA devices thereby confirming the enhanced $\xi_{DL}$ of Pt-Cr alloys. More details including formula derivation and measurement protocol can be found in the supporting information.

c. ST-FMR measurements

Besides $\xi_{DL}$, the dopant concentration dependence of Gilbert damping constant α in IMA devices is further determined via ST-FMR measurement[3], as illustrated in **Figure 5**a. By applying a microwave signal of 25 dBm of power with frequency ranging from $f = 8$ to 12 GHz onto a micron-sized strip device while sweeping an external in-plane magnetic field $H_{ext}$ at $\varphi = 45°$, magnetization precession can occur when the parameters satisfy ferromagnetic resonance (FMR) condition. This FMR response then can be read out via a rectified voltage ($V_{mix}$), which can be expressed as[3] $V_{mix} = S \times F_{sym} + A \times F_{asym}$, where $F_{sym}(H_{ext}) \equiv \Delta H^2/(\Delta H^2 + (H_{ext} - H_0)^2)$ and $F_{asym}(H_{ext}) \equiv \Delta H(H_{ext} - H_0)/(\Delta H^2 + (H_{ext} - H_0)^2)$ are symmetric and antisymmetric Lorentzian function; $S$ and $A$ are their corresponding weights; $\Delta H$ is the linewidth of the Lorentzian function; $H_0$ is the field of resonance. Representative ST-FMR field sweep results as obtained from a Pt(3)/CoFeB(2.3)/MgO(2) and a Pt$_{0.69}$Cr$_{0.31}$(3)/CoFeB(2.3)/MgO(2) device are shown in **Figure 5**b.

Next, using the $\Delta H$ versus $f$ data, damping constant α can be further extracted by

$$\Delta H = \Delta H_0 + 2\pi\alpha\frac{f}{\gamma}, \tag{5}$$

where $\Delta H_0$ is the inhomogeneous broadening; $\gamma = 1.758 \times 10^{11}$ T$^{-1}\cdot$s$^{-1}$ is the gyromagnetic ratio with g-factor ($g_e$) chosen as 2. As shown in **Figure 5**d, α is reduced from 0.078 ± 0.0006 (pure Pt) to 0.052 ± 0.0007 (x = 0.69) and shows no obvious composition dependence once Cr has been introduced into Pt. This suppressed magnetic damping is possibly attributed to the reduced magnetic proximity effect[49] or two-magnon scattering[50] in Pt/FM structures due to a weaker interfacial SOC, which is also advantageous for SOT-driven magnetization switching in in-plane magnetized devices using such Pt-based alloy.

d. SOT-driven switching measurements

To directly demonstrate the large $\xi_{DL}$ as characterized via AHE loop shift measurement and PHE curve shift measurement, current-induced SOT-driven magnetization switching is performed on the PMA devices, as depicted in **Figure 6**a. The pulse width of write/sense currents are set to be 0.05 s. For PMA devices, an in-plane magnetic field is also required to overcome $H_{DMI}$ and/or to break the switching symmetry. Representative current-induced switching loops with opposite in-plane bias fields are shown in **Figure 6**b. The switching current ($I_{sw}$) is then determined when the normalized Hall resistance ($R_H$) changes its sign, and the saturated field ($H_{sat}$) is defined as the field at which critical switching current ($I_c$) is achieved. For example, as shown in **Figure 6**c, $I_c$ is 2.10 mA and $H_{sat}$ is 1000 Oe for Pt$_{0.69}$Cr$_{0.31}$.

For a robust memory device, decent thermal stability ($\Delta = \frac{U}{k_B T}$, thermal stability factor) should also be demonstrated, where $U$ represents the energy barrier between up and down states of the FM layer; $k_B$ is the Boltzmann constant; $T$ is temperature. Since the SOT-driven switching is a thermally-assisted process, the relationship between $I_c$ and the writing current pulse width ($t_{pulse}$) follows[51]

$$I_c = I_{c0}\left[1 - \frac{1}{\Delta}\ln\left(\frac{t_{pulse}}{\tau_0}\right)\right], \tag{6}$$

where $I_{c0}$ is the zero thermal critical switching current; $\tau_0 \sim 1$ ns is the intrinsic attempt time. $\Delta$ and $I_{c0}$ therefore can be extracted from the slope and the intercept by linear fitting of $I_c$ versus $\ln\left(\frac{t_{pulse}}{\tau_0}\right)$ with varying writing current pulse-widths (100 μs < $t_{pulse}$ < 1.00 s), which is shown in **Figure 6**d.

Overall, Cr doping can significantly reduce $I_c$ due to the enhanced $\xi_{DL}$ and the reduced $H_c$. As shown in **Figure 6**e, Pt$_{0.52}$Cr$_{0.48}$ has the lowest $I_c = 0.76$ mA ($J_c \sim 9.8 \times 10^5$ A·cm$^{-2}$ in the Pt-Cr alloy layer), comparing to that of the pure Pt device: $I_c = 4.42$ mA ($J_c \sim 1.0 \times 10^7$ A·cm$^{-2}$ in the Pt layer). $H_{sat}$ of Pt-Cr series also shows the same trend of $H_{DMI}$ that Pt-Cr alloy possesses smaller spin-orbit coupling strength compared to pure Pt. As summarized in **Figure 6**f, a significant reduction of $I_{c0}$ (and $J_{c0}$ as well) is realized via Cr doping. The lowest $I_{c0} = 2.05$ mA is observed in Pt$_{0.52}$Cr$_{0.48}$-based devices ($J_{c0} \sim 2.6 \times 10^6$ A·cm$^{-2}$), compared to that of pure Pt: $I_{c0} = 10.49$ mA ($J_{c0} \sim 2.5 \times 10^7$ A·cm$^{-2}$). Moreover, for Pt$_{0.69}$Cr$_{0.31}$ that possesses the largest $\frac{H_{eff}^z}{I}$, $I_{c0} = 6.20$ mA ($J_{c0} \sim 9.3 \times 10^6$ A·cm$^{-2}$), also performs better than pure Pt. In addition, thermal stability shows no obvious degradation after a high concentration of Cr and V doping, with $\Delta = 33.58 \pm 0.47$ for Pt$_{0.69}$Cr$_{0.31}$, $\Delta = 31.12 \pm 1.30$ for the pure Pt sample from the Pt-Cr series, $\Delta = 45.45 \pm 8.13$ for Pt$_{0.80}$V$_{0.20}$, and $\Delta = 33.84 \pm 3.98$ for the pure Pt sample from the Pt-V series.

The reduction of $I_{c0}$ and the preserved $\Delta$ indicate that Pt-Cr alloy is capable of simultaneously realizing efficient SOT-driven magnetic switching and robust data retention.

## 4. Discussion

a. Spin Hall conductivity

Theoretical prediction[23] suggests that intrinsic $\sigma_{SH}$ is independent of the change of $\sigma_{xx}$ in the high $\sigma_{xx}$ regime; and $\sigma_{SH}$ would be proportional to $\sigma_{xx}^2$ ($\sigma_{SH} \propto \rho_{xx}^{-2}$) when $\rho_{xx}$ is beyond a certain limit (low $\sigma_{xx}$ regime). However, we find that there is an additional scattering contribution from the Cr dopants within the conductive metal regime, as shown in **Figure 7**a. Comparing to the pure Pt case where $\sigma_{SH}^{Pt}$ is $4.47 \times 10^5 \left(\frac{\hbar}{2e}\right) \Omega^{-1} \cdot m^{-1}$, the largest $\sigma_{SH}^{Pt-Cr}$ is $6.86 \times 10^5 \left(\frac{\hbar}{2e}\right) \Omega^{-1} \cdot m^{-1}$ is observed for Pt$_{0.75}$Cr$_{0.25}$ in this work. We firstly conclude that the interfacial spin-orbit coupling (iSOC) contributes little to the overall SHC enhancement[52, 53], based on that their opposite trend (as evidenced by the reduced DMI effective field). One of the possible origins of this enhancement of SHC might be related to the weak ferromagnetism of Pt-Cr alloy[54]. The magnetic impurities can act as a spin-dependent barrier and results in an additional SHC. However, it requires more works to unveil its true origin (evidence of weak ferromagnetism in our sputtered Pt-Cr is shown in supporting information). Similar results of enhanced $\sigma_{SH}$ through Cr alloying has also been reported in the W-Cr/YIG system[55]. Another possible origin of the SHC enhancement is the OHE from the Cr content. It is possible that the sizable orbital current from Cr converts into spin current at the presence of the strong SOC from Pt. Therefore, the observed giant SHC consists of the contributions from the SHE of the Pt content and the OHE of the Cr content. Besides the SHC enhancement, the decay of $\sigma_{SH}$ is also mitigated in the resistive regime ($\sigma_{SH}^{Pt-C} \propto \sigma_{xx}^{0.9}$). Combining the $\sigma_{SH}$ enhancement in the conductive regime and the suppressed reduction in the resistive regime, $\sigma_{SH}$ of resistive Pt-Cr can remain almost the same as that of pure Pt even with $\rho_{xx} \sim 183.9$ μΩ·cm. This characteristic brings versatility to employ Pt alloys in a wide range of $\rho_{xx}$ for SOT-related as well as other spintronics applications.

b.  Power consumption benchmark

The spin-torque switching efficiency ($\varepsilon$) is defined as $\varepsilon \equiv \frac{\Delta}{I_{c0}}$, which can be considered as one of the key figures of merits to discuss SOT switching performance. As summarized in **Figure 7b**, the reduced $I_{c0}$ together with a fairly unchanged $\Delta$ lead to the improvement on $\varepsilon$, which is $5.41 \pm 0.05$ mA$^{-1}$ for Pt$_{0.69}$Cr$_{0.31}$ as compared to $2.96 \pm 0.09$ mA$^{-1}$ for pure Pt.

Power consumption per volume without current shunting and thermal effect is estimated by $p_0 = \rho_{xx} J_{c0}^2$. $p_0$ of the pure Pt sample is about thirty times larger than the Pt$_{0.52}$Cr$_{0.48}$ case. The lowest $p_0$ is estimated to be as low as $1.03 \times 10^{12}$ mW·cm$^{-3}$ for Pt$_{0.52}$Cr$_{0.48}$. A fairly small $J_{c0}$ of Pt$_{0.52}$Cr$_{0.48}$ results in a significant reduction of power consumption even with high $\rho_{xx}$. We further provide a comprehensive comparison on power consumption based on a prototypic SOT-MRAM device structure with various SCSs: Pt-Cr, Pt-V, Pt-based[12-16, 56, 57], β-Ta[7], β-W[37] and chalcogenide-based[58-60] SCSs. The stack is chosen as SCS(5)/CoFeB(1) with $\rho_{CoFeB} = 134$ μΩ·cm. The device geometry is $w = 2$ μm, $l_{FM} = 2$ μm, $l_{SCS} = 3$ μm. The magnetic properties of the perpendicular magnetized CoFeB layer are set as $M_s = 1000$ emu·cm$^{-3}$ and $H_c = 100$ Oe. The calculated results show that Pt$_{0.69}$Cr$_{0.31}$ provides a comparably low power consumption among Pt-based SCSs due to its large efficiency and moderate $\rho_{xx}$, as shown in **Figure 7c** and **Table 1** (details of calculations in the supporting information). It is noteworthy that we use the DL-SOT efficiencies based on the nominal width rather than the corrected effective width for a fairer comparison to those prior works used Hall bar devices as well.

## 5. Conclusion

In this work we report a simple alloying approach to significantly enhance $\xi_{DL}$ of Pt-based heterostructures by tuning $\rho_{xx}$. It is found that both V and Cr dopants have such effect, where Pt$_{0.69}$Cr$_{0.31}$ possesses the largest DL-SOT efficiency of $\xi_{DL} \sim 0.9$ with $\rho_{xx}^{Pt}$ $\sim 130$ μΩ·cm (average $\sigma_{SH} \sim 6.5 \times 10^5 \left(\frac{\hbar}{2e}\right) \Omega^{-1} \cdot m^{-1}$) among all tested samples. We verify this high efficiency from both Pt-Cr/Co (PMA) and Pt-Cr/CoFeB (IMA) samples through AHE loop shift and PHE curve shift measurements, respectively. ST-FMR measurement further shows that damping constant α can be reduced from 0.078 to 0.052 via Cr doping. A low critical switching current density of $J_c \sim 9.8 \times 10^5$ A·cm$^{-2}$ using Pt$_{0.52}$Cr$_{0.48}$ is also achieved. These results suggest that employing the large SHC Pt, a classical spin Hall metal, with optimized 3$d$ element alloying is the key to realize giant DL-SOT efficiency towards or even beyond unity.

**ASSOCIATED CONTENT**

**Supporting Information**

Uncertainties estimation by error propagation; methods of alloy preparation; magnetic characterization of Co and CoFeB layer; formula derivation of planar Hall effect curve shift measurement.

**Experimental section**

Magnetic properties are characterized by vibrating sample magnetometer (MicroSense VSM EZ9) and superconducting quantum interference device (Quantum Design MPMS 3(DC)). DL-SOT efficiency characterization and SOT-driven switching measurement for PMA samples are performed with a DC current source (Keithley 2400 or Keysight Agilent 8110A) and a voltage meter (Keithley 2000). DL-SOT characterization for IMA samples are performed using a vector field magnet (GMW Associates 5204), a DC current source (Keithley 2400) and a voltage meter (Keithley 2000). ST-FMR measurements for IMA samples are performed with a signal generator (Keysight E8257D) to generate the modulated RF current, and a lock-in amplifier (Signal Recovery 7265) for signal detection.

**Acknowledgements**

This work is supported by the Ministry of Science and Technology of Taiwan (MOST) under grant No. MOST 109-2636-M-002-006 and by the Center of Atomic Initiative for New Materials (AI-Mat), National Taiwan University from the Featured Areas Research Center Program within the framework of the Higher Education Sprout Project by the Ministry of Education (MOE) in Taiwan under grant No. NTU-108L9008. This work is also partly supported by Taiwan Semiconductor Manufacturing Company (TSMC).

**Conflict of Interest**

All authors have no conflicts of interest.

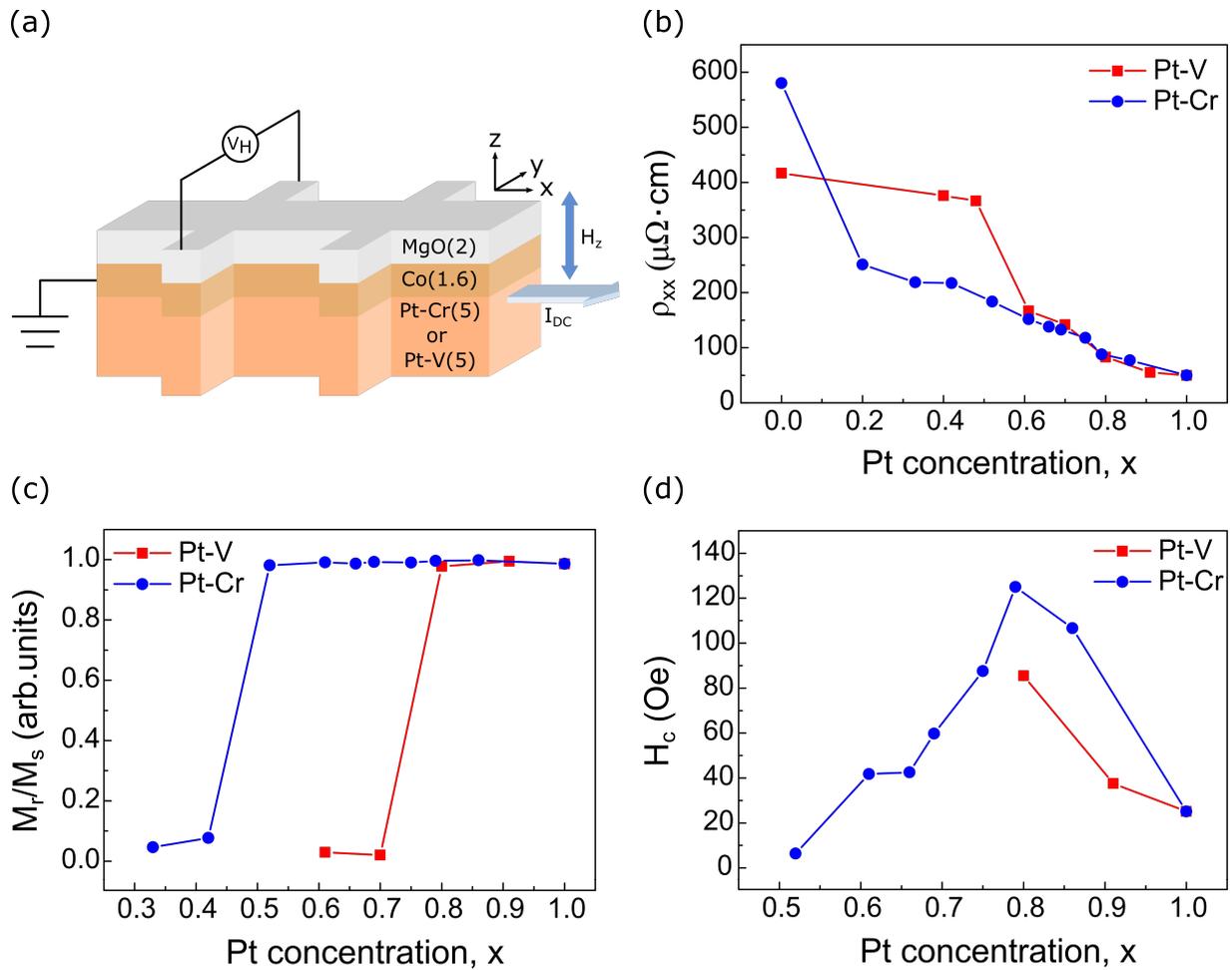

**Figure 1.** Experimental setup and materials information of $Pt_xCr_{1-x}(5)/Co(1.6)/MgO(2)$ and $Pt_xV_{1-x}(5)/Co(1.6)/MgO(2)$ magnetic heterostructures with varying Pt concentration (denoted as x). (a) Schematics of Hall bar devices, including their structures and circuits setup for Hall voltage measurements. (b) Longitudinal resistivities of $Pt_xCr_{1-x}(5)$ and $Pt_xV_{1-x}(5)$ (c) Ratio of remnant magnetization to saturation magnetization and (d) out-of-plane coercivity of $Pt_xCr_{1-x}(5)/Co(1.6)/MgO(2)$ and $Pt_xV_{1-x}(5)/Co(1.6)/MgO(2)$ heterostructures.

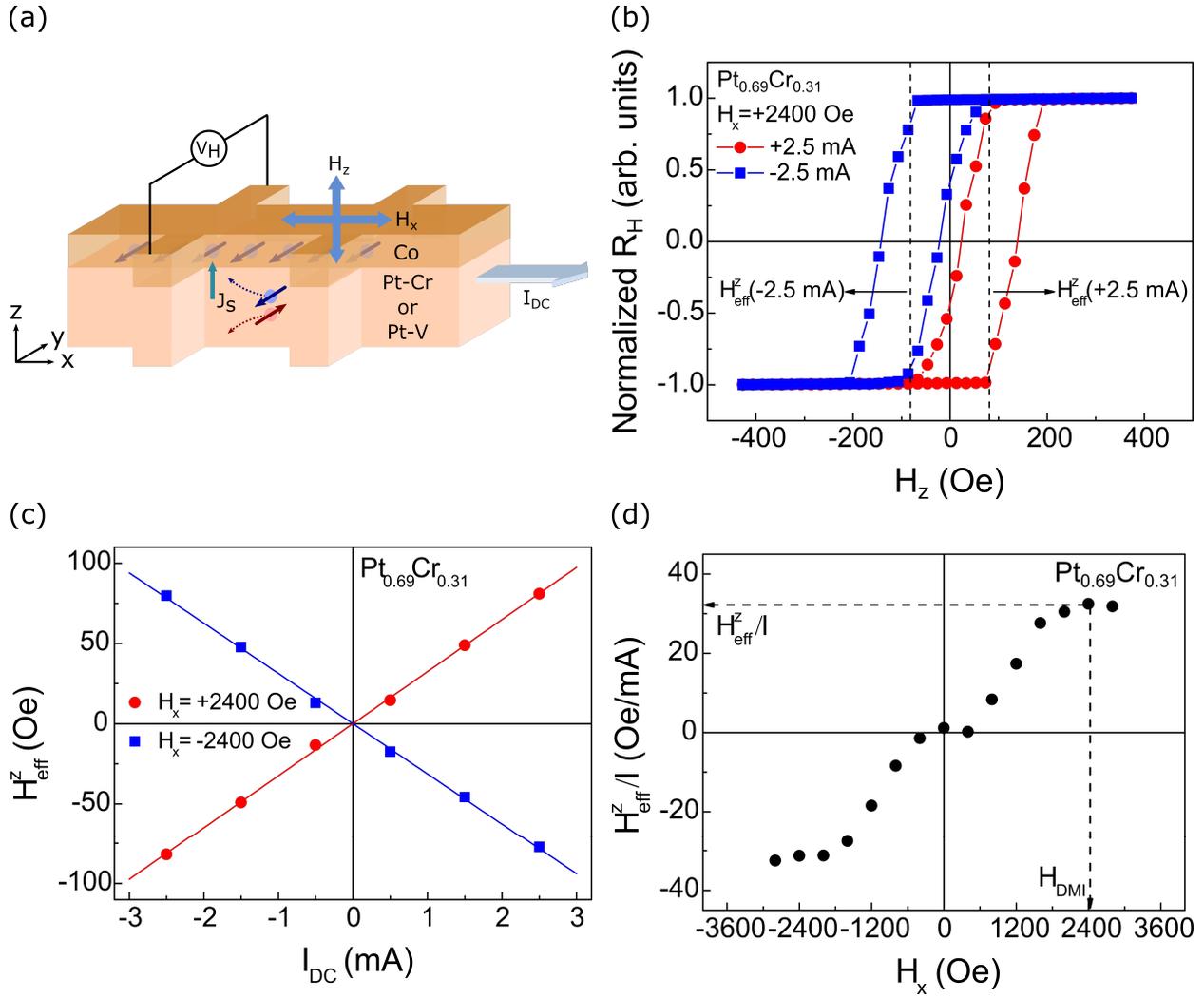

**Figure 2.** Anomalous Hall hysteresis loop-shift measurements. All representative results are obtained from $Pt_{0.69}Cr_{0.31}$-based PMA devices. (a) Schematics of experimental setup, where $I_{DC}$ is the applied in-plane current along $x$-direction; $H_x$ and $H_z$ are the applied external magnetic fields. (b) Representative results of hysteresis loop-shift measurements under $H_x = 2400$ Oe and $I_{DC} = \pm 2.5$ mA. (c) Representative linear relationship between $H_{eff}^z$ and $I_{DC}$ under $H_x = \pm 2400$ Oe. (d) $H_{eff}^z/I$ versus $H_x$. The pointed-out data correspond to the saturated $H_{eff}^z/I$ and $H_{DMI}$.

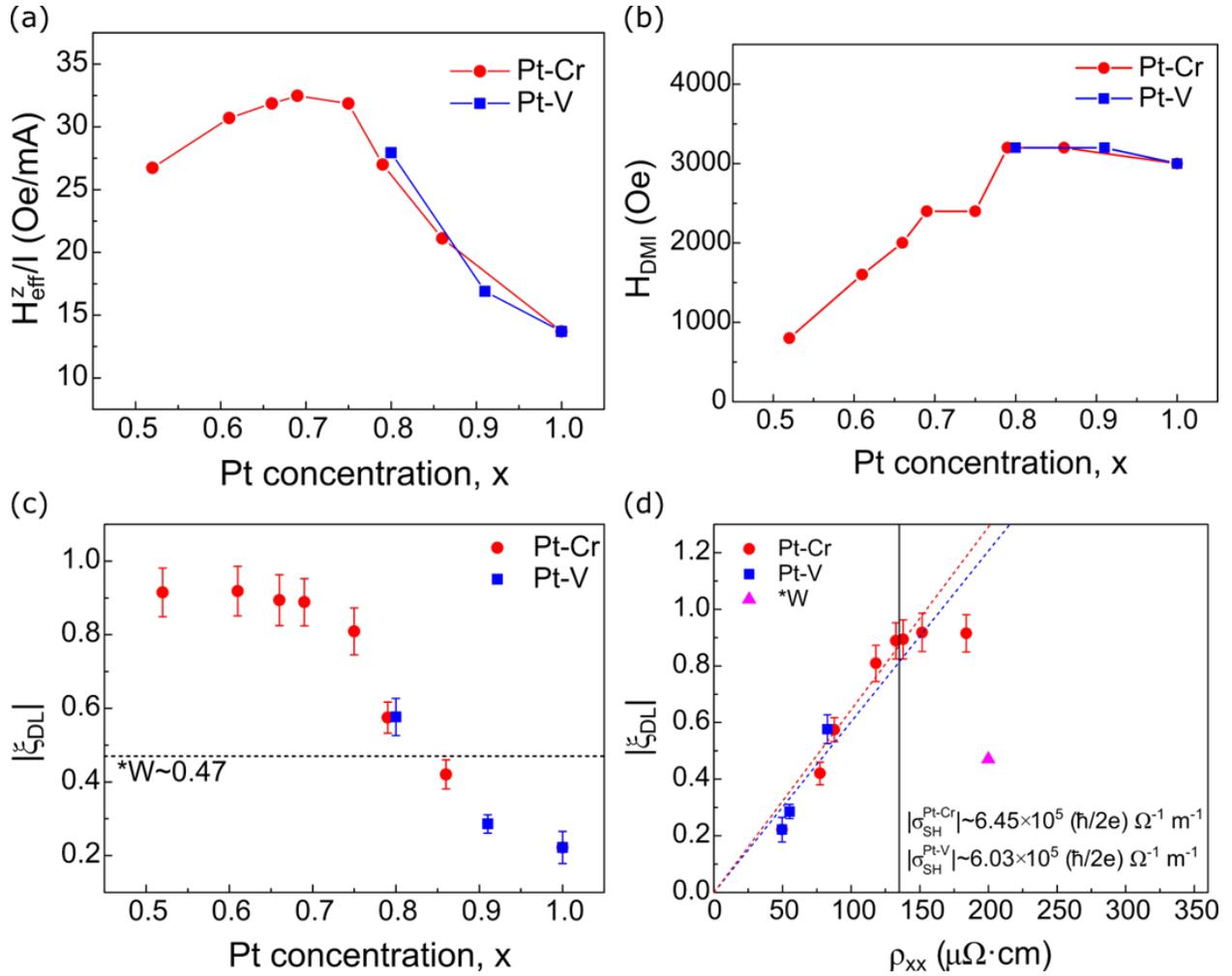

**Figure 3.** DL-SOT efficiency ($\xi_{DL}$) of $Pt_xCr_{1-x}$ and $Pt_xV_{1-x}$ in PMA devices. (a~c) $H_{eff}^z/I$, $H_{DMI}$ and the estimated $\xi_{DL}$ as functions of Pt concentration. (d) $\xi_{DL}$ as a function of the SCS layer resistivity. The compared data of W is from Y. Takeuchi, *et al.*[37] The dashed lines represent linear fits to the data. Note that for the Pt-Cr case the fitting range is from $\rho_{xx} = 0$ to $135\ \mu\Omega \cdot cm$.

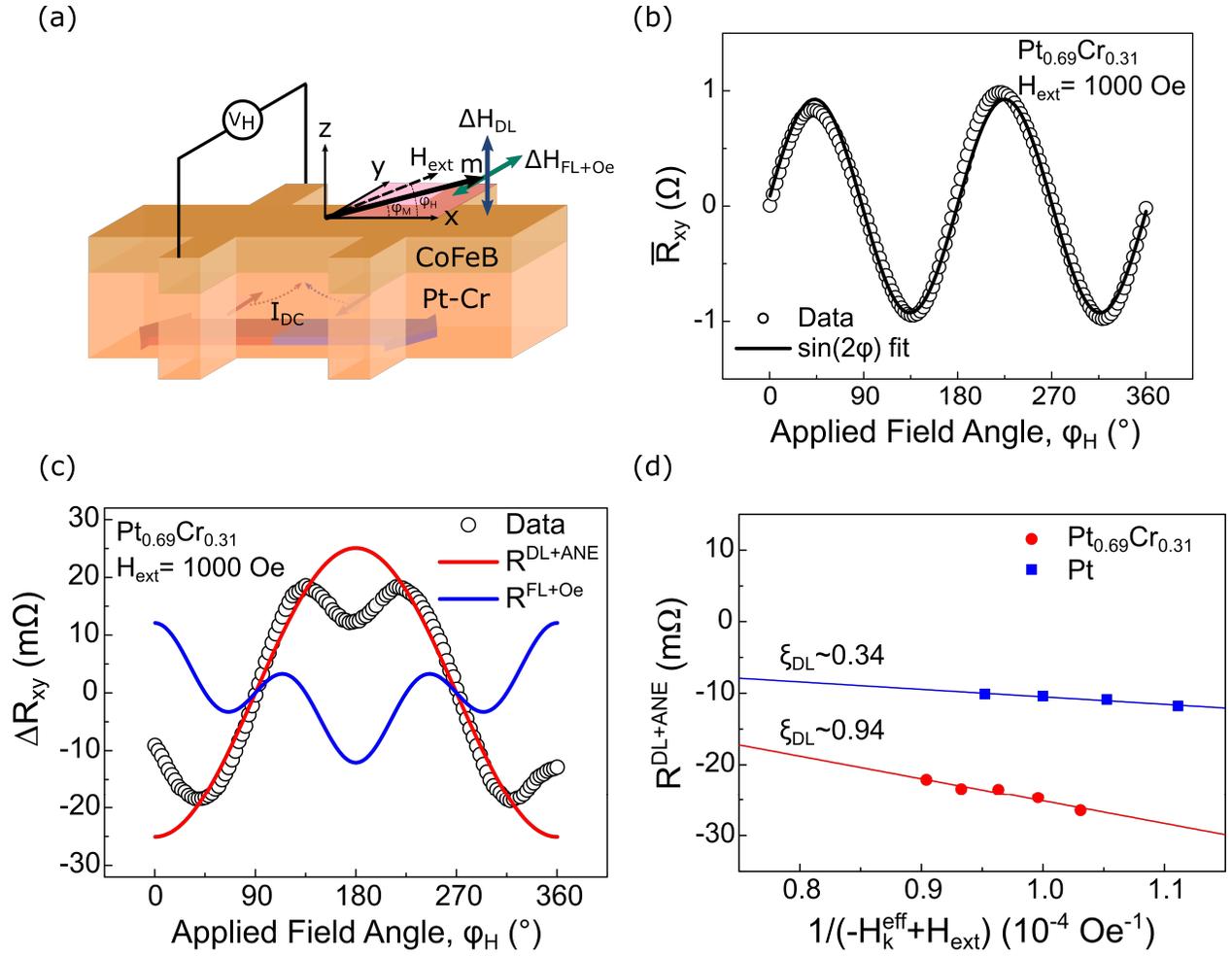

**Figure 4.** Planar Hall curve-shift measurements. All representative results are from $Pt_{0.69}Cr_{0.31}$-based IMA devices. (a) Schematics of experimental setup, where $I_{DC}$ is the applied in-plane current along $x$-direction; $H_{ext}$ is the applied field within $x$-$y$ plane; $\Delta H_{DL,FL,Oe}$ respectively are current-induced DL-SOT field, FL-SOT field and Oersted field; $m$ is the in-plane magnetic vector; $\varphi_M$ and $\varphi_H$ respectively are the azimuthal angle of $m$ and $H_{ext}$. (b, c) Representative $\overline{R_{xy}}$ and $\Delta R_{xy}$ under $H_{ext} = 1000$ Oe and $I_{DC} = \pm 3$ mA. (d) Linear fittings on the extracted $R^{DL}$ versus $\frac{1}{-H_k^{eff}+H_{ext}}$.

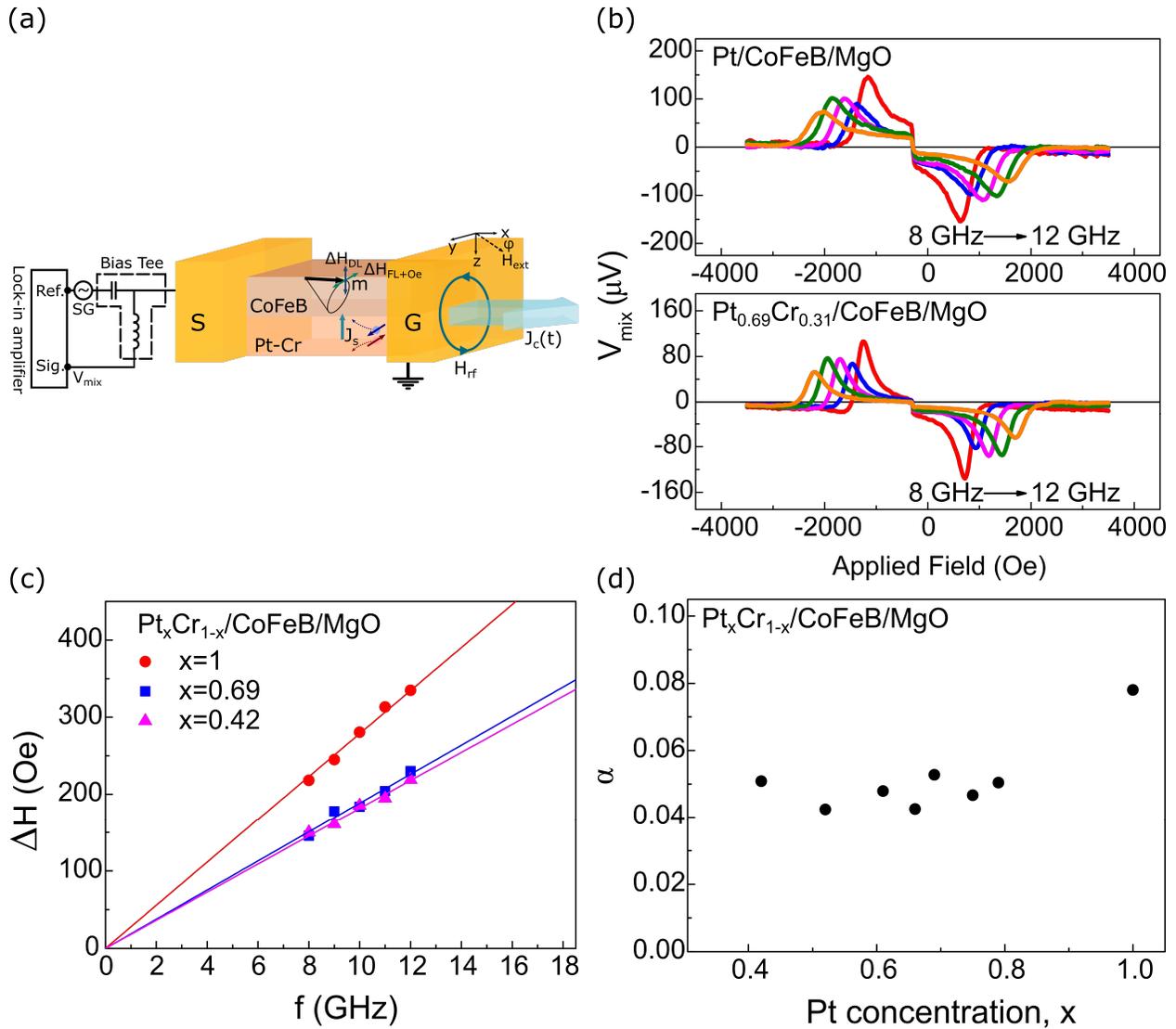

**Figure 5.** Gilbert damping constant α of IMA devices characterized via ST-FMR. (a) Schematics of ST-FMR measurement. The rectified ST-FMR signal ($V_{\text{mix}}$) is extracted through a bias-tee and by a lock-in amplifier. (b) Representative field sweep ST-FMR signals from pure-Pt and $Pt_{0.69}Cr_{0.31}$-based IMA devices. (c) The microwave frequency ($f$) dependence of FMR linewidth ($\Delta H$) for pure Pt and two representative Pt-Cr devices. (d) The extracted α versus Pt concentration.

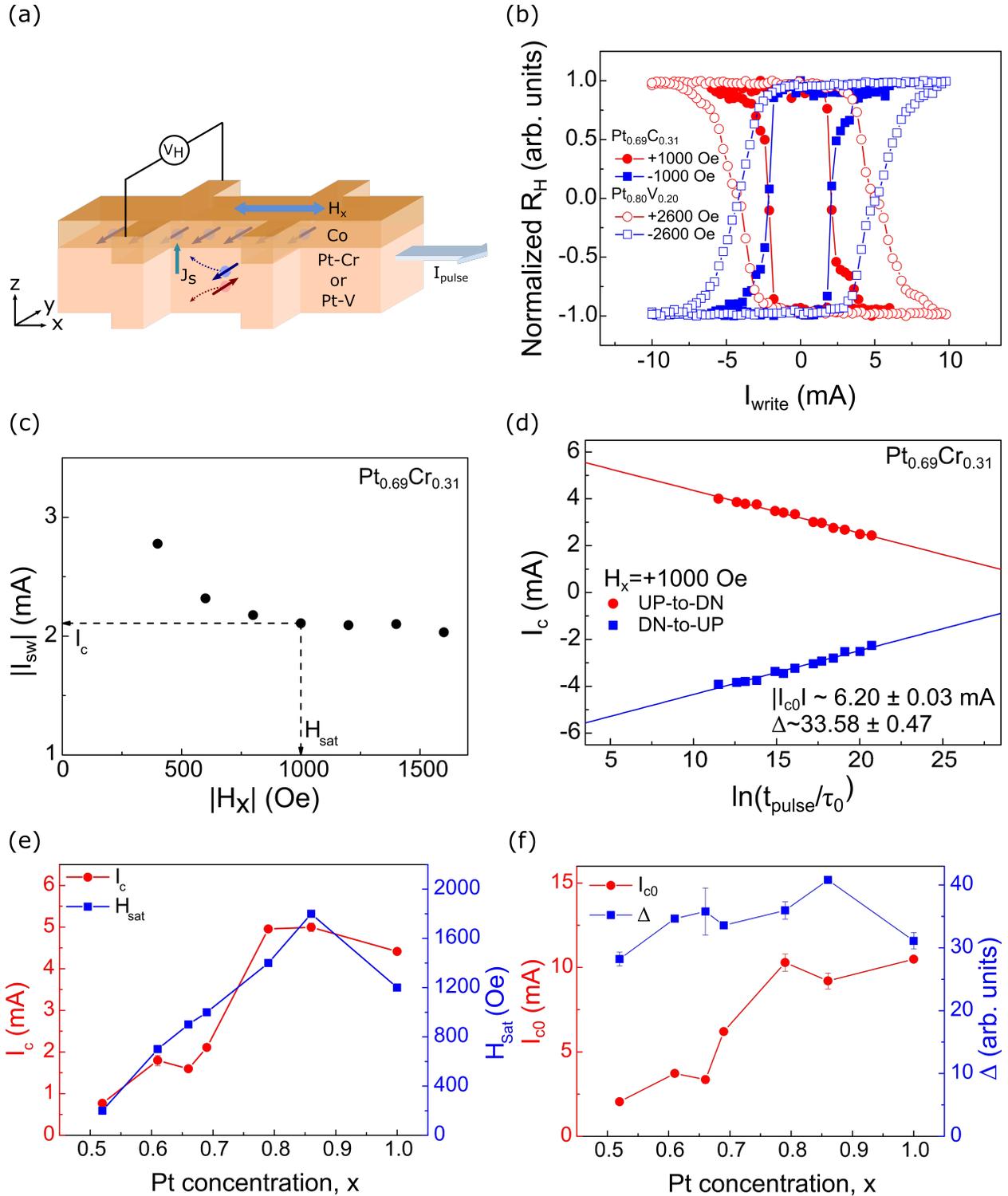

**Figure 6.** Current-induced SOT switching. The representative results are from Pt$_{0.69}$Cr$_{0.31}$(Pt$_{0.80}$V$_{0.20}$)-based PMA devices. (a) Schematics of experimental setup. Both the applied current pulse ($I_{pulse}$) and the applied field ($H_x$) are the along $x$-direction; (b) Representative SOT

switching under $H_\text{x} = \pm1000$ Oe for a $Pt_{0.69}Cr_{0.31}$ based device, and under $H_\text{x} = \pm2600$ Oe for a $Pt_{0.8}V_{0.2}$-based device. (c) Switching current ($|I_\text{sw}|$) as the function of $|H_x|$, the critical switching current ($|I_\text{c}|$) is the saturated $|I_\text{sw}|$ with corresponding saturation field ($H_\text{sat}$). (d) $I_\text{c}$ versus different pulse width ($t_\text{pulse}$) under $H_\text{x} = 1000$ Oe. (e, f) The summary of $I_\text{c}$, $H_\text{sat}$ and extracted $I_\text{c0}$, thermal stability ($\Delta$) as functions of Pt concentration.

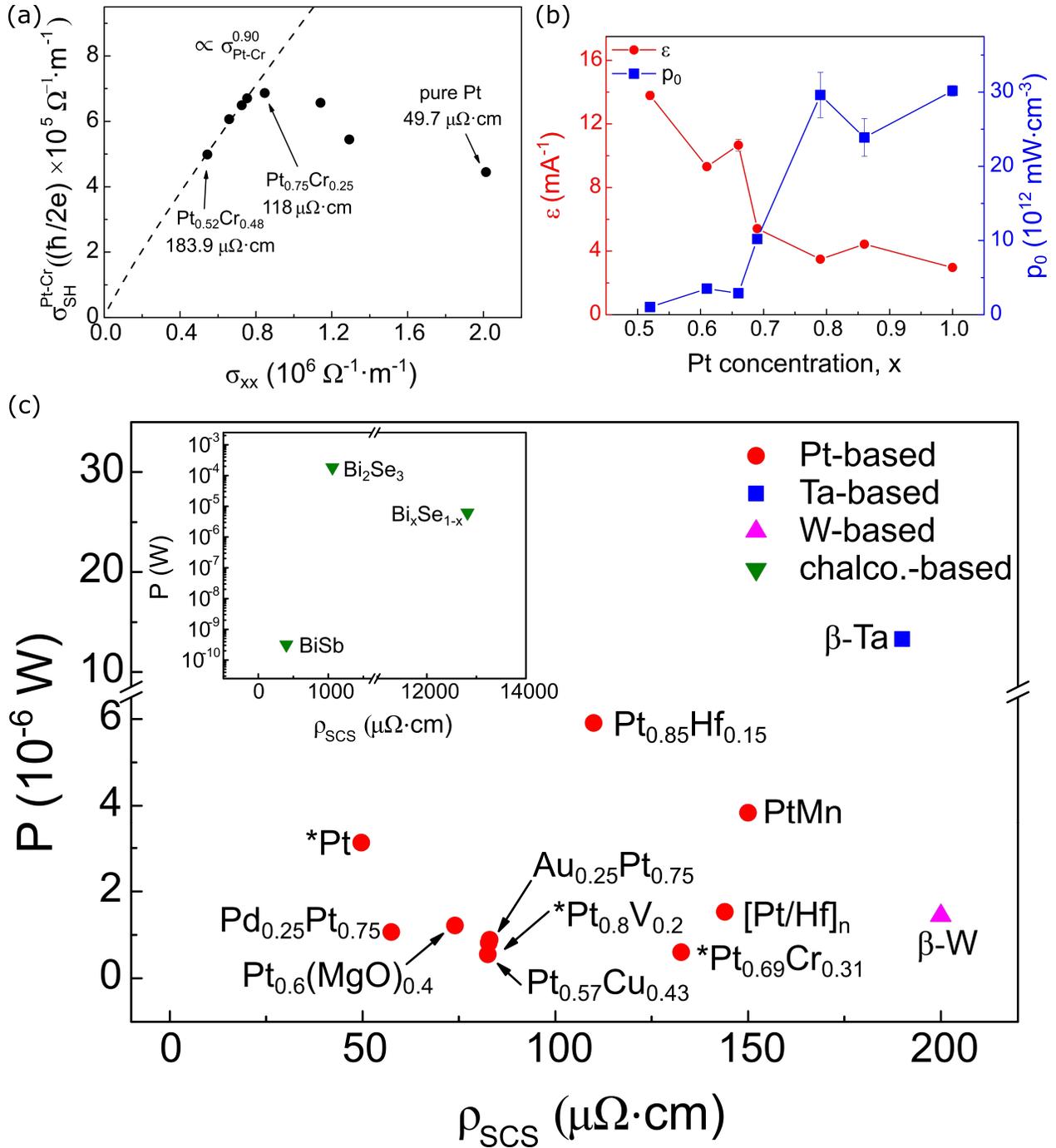

**Figure 7.** Benchmarking SOT switching performance of Pt-Cr alloy. (a) Spin Hall conductivity ($\sigma_{SH}^{Pt-C}$) versus longitudinal conductivity ($\sigma_{xx}^{Pt-C}$) of Pt-Cr alloy. (b) Switching efficiency ($\varepsilon$) and power consumption without current shunting and thermal effect ($p_0$) as functions of Pt

concentration. (c) Summary of calculated power consumption ($P$) versus longitudinal resistivities of reported SCS ($\rho_{SCS}$). The starred points represent the results from this work.

| SCS | $\rho_{SCS}$ [$\mu\Omega\cdot$cm] | $\xi_{DL}$ [1] | $J_c^{SCS}$ [A$\cdot$cm$^{-2}$] | $P$ [W] | Technique | Device | Ref. |
|---|---|---|---|---|---|---|---|
| Pt-based | | | | | | | |
| Pt-Cr | 132.8 | 0.56 | $3.43 \times 10^6$ | $5.60 \times 10^{-7}$ | Loop shift | Hall bar | This work |
| Pt-V | 82.8 | 0.36 | $5.29 \times 10^6$ | $7.78 \times 10^{-7}$ | Loop shift | Hall bar | This work |
| Pt | 49.7 | 0.14 | $1.40 \times 10^7$ | $3.12 \times 10^{-6}$ | Loop shift | Hall bar | This work |
| $Pt_{0.57}Cu_{0.43}$ | 82.5 | 0.44 | $4.36 \times 10^6$ | $5.51 \times 10^{-7}$ | Loop shift | Hall bar | [16] |
| PtMn | 150 | 0.24 | $8.04 \times 10^6$ | $3.83 \times 10^{-6}$ | Harmonic | Hall bar | [56] |
| $Au_{0.25}Pt_{0.75}$ | 83 | 0.35 | $5.52 \times 10^6$ | $8.87 \times 10^{-7}$ | Harmonic | Hall bar | [12] |
| $Pd_{0.25}Pt_{0.75}$ | 57.5 | 0.26 | $7.43 \times 10^6$ | $1.06 \times 10^{-6}$ | Harmonic | Hall bar | [13] |
| $Pt_{0.6}(MgO)_{0.4}$ | 74 | 0.28 | $6.90 \times 10^6$ | $1.22 \times 10^{-6}$ | Harmonic | Hall bar | [14] |
| $Pt_{0.85}Hf_{0.15}$ | 110 | 0.16 | $1.21 \times 10^7$ | $5.90 \times 10^{-6}$ | Harmonic | Hall bar | [15] |
| $[Pt/Hf]_n$ | 144 | 0.37 | $5.22 \times 10^6$ | $1.53 \times 10^{-6}$ | Harmonic | Hall bar | [57] |
| Ta-based | | | | | | | |
| $\beta$-Ta | 190 | 0.15 | $1.29 \times 10^7$ | $1.33 \times 10^{-5}$ | ST-FMR | Stripe | [7] |
| W-based | | | | | | | |
| $\beta$-W | 200 | 0.47 | $4.11 \times 10^6$ | $1.45 \times 10^{-6}$ | Harmonic | Hall bar | [37] |
| chalco.-based | | | | | | | |
| $Bi_2Se_3$ | 1060 | 0.16 | $1.21 \times 10^7$ | $1.83 \times 10^{-4}$ | Loop shift | Hall bar | [58] |
| $Bi_xSe_{1-x}$ | 12821 | 18.6 | $1.04 \times 10^5$ | $6.15 \times 10^{-6}$ | PHE-DC | Hall bar | [59] |
| BiSb | 400 | 52 | $3.71 \times 10^4$ | $3.17 \times 10^{-10}$ | AHE-DC | Hall bar | [60] |

**Table 1.** Summary of $\rho_{SCS}$, $\xi_{DL}$, corresponding estimated $J_c$ and $P$ of reported spin current sources (SCSs) based on an ideal, prototypic SOT-MRAM device with PMA. Loop shift, Harmonic, and (AHE) PHE-DC are abbreviation of AHE hysteresis loop-shift measurement, harmonic Hall voltage measurement, and (AHE) PHE DC-induced curve-shift measurement, respectively.

**Table of Content**

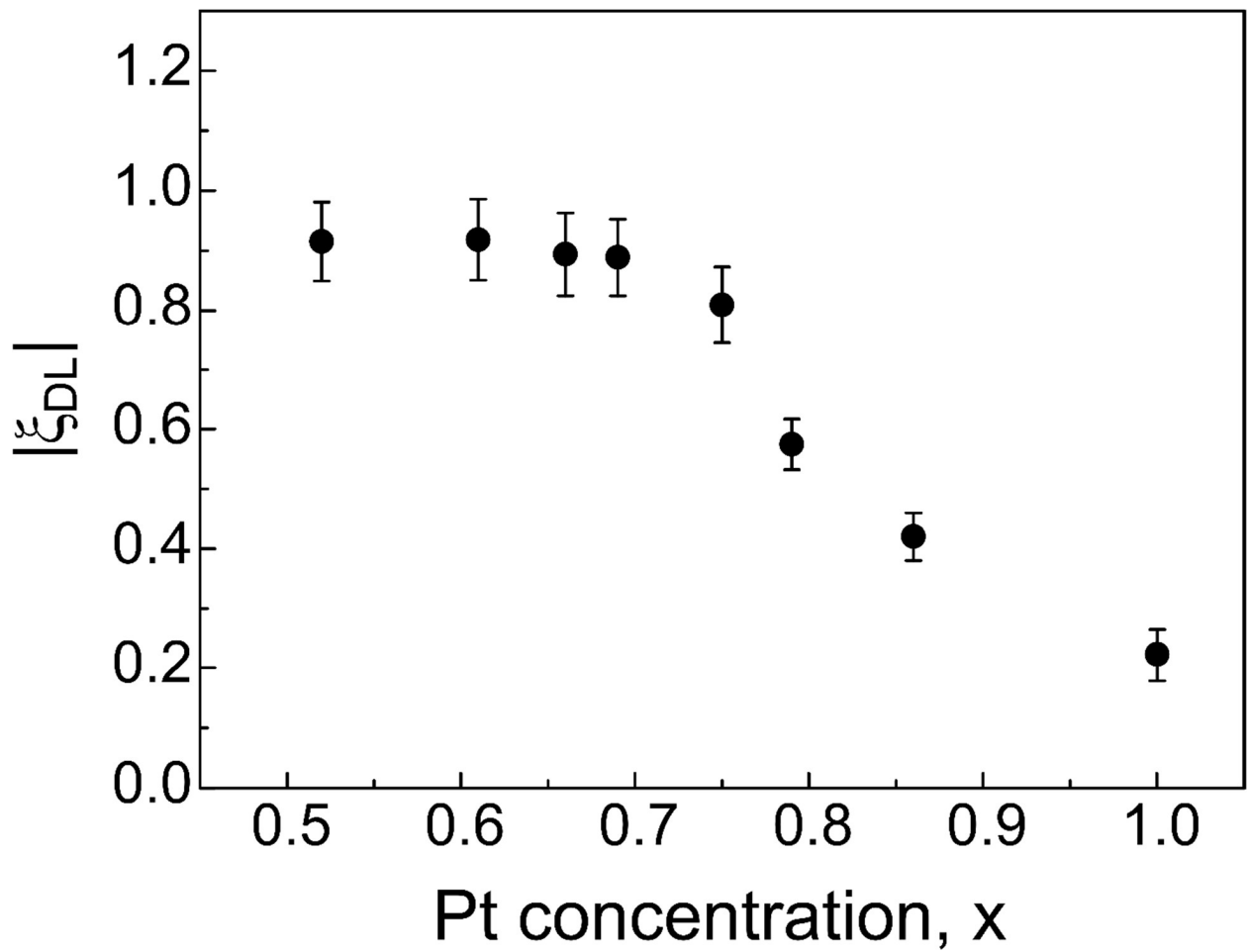

# Supporting information

**Toward 100% Spin-Orbit Torque Efficiency with High Spin-Orbital Hall Conductivity Pt-Cr Alloys**

*Chen-Yu Hu[†], Yu-Fang Chiu[†], Chia-Chin Tsai, Chao-Chung Huang, Kuan-Hao Chen, Cheng-Wei Peng, Chien-Min Lee, Ming-Yuan Song, Yen-Lin Huang, Shy-Jay Lin, and Chi-Feng Pai\**

Chen-Yu Hu, Yu-Fang Chiu, Chia-Chin Tsai, Chao-Chung Huang, Kuan-Hao Chen, Prof. Chi-Feng Pai

Department of Materials Science and Engineering, National Taiwan University

Taipei 10617, Taiwan

E-mail: cfpai@ntu.edu.tw

Chien-Min Lee, Ming-Yuan Song, Yen-Lin Huang, Shy-Jay Lin

Corporate Research, Taiwan Semiconductor Manufacturing Company

Hsinchu 30078, Taiwan

1. **Estimation on the uncertainties of characterized physical parameters**

The resulting uncertainties comes from the error propagation from each physical parameter that we characterized. We first define the notation for the following parameters. For a measured physical quantity, $M$, it can be represented as

$$M = \bar{M} + e_M, \tag{S1}$$

where $\bar{M}$ is the mean, and $e_M$ is the estimation on error (typically from the standard deviation, STD). Take the switching efficiency ($\varepsilon \equiv \frac{\Delta}{I_{c0}}$) for instance. From the linear fitting on various pulse width versus critical switching current, we can get $\Delta = \bar{\Delta} + e_\Delta$ and $I_{c0} = \overline{I_{c0}} + e_{I_{c0}}$ either from the extracted slope or the intercept. Then we can derive the relation between the resulting quantities and the measured physical quantities. The detailed derivation is as the following,

$$\varepsilon \equiv \frac{\Delta}{I_{c0}} = \frac{\bar{\Delta} + e_\Delta}{\overline{I_{c0}} + e_{I_{c0}}} = \frac{\bar{\Delta} + e_\Delta}{\overline{I_{c0}}} \frac{1}{1 + \frac{e_{I_{c0}}}{\overline{I_{c0}}}}$$

$$\xrightarrow{\text{Taylor expansion}} \frac{\bar{\Delta} + e_\Delta}{\overline{I_{c0}}} \left(1 - \frac{e_{I_{c0}}}{\overline{I_{c0}}} + \left(\frac{e_{I_{c0}}}{\overline{I_{c0}}}\right)^2 + \cdots\right) \sim \frac{\bar{\Delta} + e_\Delta}{\overline{I_{c0}}} \left(1 - \frac{e_{I_{c0}}}{\overline{I_{c0}}}\right)$$

$$\varepsilon \sim \left(\frac{\bar{\Delta}}{\overline{I_{c0}}} + \frac{e_\Delta}{\overline{I_{c0}}}\right)\left(1 - \frac{e_{I_{c0}}}{\overline{I_{c0}}}\right) = \left(\bar{\varepsilon} + \frac{e_\Delta}{\overline{I_{c0}}}\right)\left(1 - \frac{e_{I_{c0}}}{\overline{I_{c0}}}\right) = \bar{\varepsilon} + \frac{e_\Delta - \bar{\varepsilon} e_{I_{c0}}}{\overline{I_{c0}}} - \frac{e_\Delta e_{I_{c0}}}{(\overline{I_{c0}})^2} \tag{S2}$$

$$= \bar{\varepsilon} + e_\varepsilon$$

where

$$\bar{\varepsilon} \equiv \frac{\bar{\Delta}}{\overline{I_{c0}}}$$

$$e_\varepsilon \equiv \frac{e_\Delta - \bar{\varepsilon} e_{I_{c0}}}{\overline{I_{c0}}} - \frac{e_\Delta e_{I_{c0}}}{(\overline{I_{c0}})^2}.$$

Other uncertainties mentioned in the manuscript are estimated based on the same method.

## 2. Methods of Pt-Cr and Pt-V alloy preparation

We first separately calibrated the deposition rate ($r_{thickness}$) of Pt, V and Cr sputter target in our magnetron sputtering system under various sputtering power. $r_{thickness}$ is then estimated by the measured film thickness divided growth time, where the thickness is characterized by atomic force microscope (AFM). And the molar deposition rate ($r_{molar}$) is proportional to the following equation,

$$r_{molar} \propto \frac{d \times r_{thickness}}{m}, \quad (S3)$$

where $d$ and $m$ respectively are the density and the atomic weight of the deposited materials. Here we choose $d_{Pt} = 21.45 \text{ g} \cdot \text{cm}^{-3}$; $d_{Cr} = 7.19 \text{ g} \cdot \text{cm}^{-3}$; $d_V = 6.11 \text{ g} \cdot \text{cm}^{-3}$; $m_{Pt} = 195.084 \text{ u}$, $m_{Cr} = 51.996\ u$ and $m_V = 50.6942\ u$. Therefore, the alloying composition prepared by co-sputtering can be determined by the ratio of each molar deposition rate. For example, we seperately characterized that the $r^{Pt}_{thickness} \sim 0.0900 \text{ nm} \cdot \text{s}^{-1}$ under 30 W, and $r^{Cr}_{thickness} \sim 0.0467 \text{ nm} \cdot \text{s}^{-1}$ under 30 W. Then the Pt concentration of $Pt_xCr_{1-x}$ under co-sputtering is determined as $\frac{r^{Pt}_{molar}}{r^{Pt}_{molar}+r^{Cr}_{molar}} \sim 0.61$, that's the reason why we state the Pt-Cr alloy under this preparation method being $Pt_{0.61}Cr_{0.31}$. The following tables are the preparation details of Pt-Cr and Pt-V alloys. The difference of Pt sputtering rate is the result from different preparation date.

| Power (W) | | sputtering rate (nm/s) | | | molar sputtering rate (#/s) | | concentration | |
|---|---|---|---|---|---|---|---|---|
| Pt | Cr | Pt | Cr | Total | Pt | Cr | Pt | Cr |
| 50 | 10 | 0.15 | 0.01917 | 0.16917 | 0.01649 | 0.00265 | 0.86 | 0.14 |
| 40 | 30 | 0.13 | 0.04667 | 0.17667 | 0.01429 | 0.00645 | 0.69 | 0.31 |
| 30 | 50 | 0.09 | 0.065 | 0.155 | 0.0099 | 0.00899 | 0.52 | 0.48 |
| 30 | 30 | 0.09 | 0.04667 | 0.13667 | 0.0099 | 0.00645 | 0.61 | 0.39 |
| 30 | 10 | 0.09 | 0.01917 | 0.10917 | 0.0099 | 0.00265 | 0.79 | 0.21 |
| 30 | 0 | 0.09 | 0 | 0.09 | 0.0099 | 0 | 1.00 | 0.00 |
| 20 | 50 | 0.06 | 0.065 | 0.125 | 0.0066 | 0.00899 | 0.42 | 0.58 |
| 10 | 40 | 0.03333 | 0.05333 | 0.08667 | 0.00367 | 0.00737 | 0.33 | 0.67 |
| 5 | 50 | 0.02 | 0.065 | 0.085 | 0.0022 | 0.00899 | 0.20 | 0.80 |
| 0 | 30 | 0 | 0.04667 | 0.04667 | 0 | 0.00645 | 0.00 | 1.00 |

**Table S1.** The details of Pt-Cr alloy preparation, including the sputtering power of each target, the corresponding (atomic) sputtering rate and the resulting alloy concentration.

| Power (W) | | sputtering rate (nm/s) | | | molar sputtering rate (#/s) | | concentration | |
|---|---|---|---|---|---|---|---|---|
| Pt | V | Pt | V | Total | Pt | V | Pt | V |
| 40 | 30 | 0.11 | 0.025 | 0.135 | 0.01209479 | 0.002999 | 0.80 | 0.20 |
| 40 | 10 | 0.11 | 0.01 | 0.12 | 0.01209479 | 0.001199 | 0.91 | 0.09 |
| 30 | 50 | 0.09 | 0.035 | 0.125 | 0.00989574 | 0.004198 | 0.70 | 0.30 |
| 30 | 0 | 0.09 | 0 | 0.09 | 0.00989574 | 0 | 1.00 | 0.00 |
| 20 | 50 | 0.06 | 0.035 | 0.095 | 0.00659716 | 0.004198 | 0.61 | 0.39 |
| 5 | 50 | 0.025 | 0.035 | 0.06 | 0.00274882 | 0.004198 | 0.40 | 0.60 |
| 5 | 30 | 0.025 | 0.025 | 0.05 | 0.00274882 | 0.002999 | 0.48 | 0.52 |
| 0 | 30 | 0 | 0.025 | 0.025 | 0 | 0.002999 | 0.00 | 1.00 |

**Table S2.** The details of Pt-V alloy preparation, including the sputtering power of each target, the corresponding (atomic) sputtering rate and the resulting alloy concentration.

### 3. Magnetic properties characterization

The saturation magnetization ($M_s$) and the magnetic dead layer thickness ($t_{\text{dead}}$) of our magnetic heterostructures are characterized via vibrating sample magnetometer (VSM) and superconducting quantum interference device (SQUID), respectively. The $M_s$ and $t_{\text{dead}}$ can be extracted by the linear fitting of $\frac{M}{A}$ versus $t_{\text{FM}}$ since

$$\frac{M}{A} = M_s \times (t_{FM} - t_{dead}), \tag{S4}$$

where $M$ is the measured magnetization; the exact sample area ($A$) is 0.25 cm$^2$; $t_{FM}$ is the thickness of the deposited magnetic layer.

### 3.1. Pt$_x$Cr$_{1-x}$/Co/MgO structure

The $M_s$ and $t_{dead}$ of Pt(5)/Co($t_{Co}$)/MgO(2) capped by Ta(6) are $M_s = 1049$ emu/cm$^3$, $t_{dead} = 0.36$ nm by VSM; $M_s = 1182$ emu/cm$^3$, $t_{dead} = 0.61$ nm by SQUID, giving an average $M_s = 1116 \pm 94$ emu/cm$^3$ and $t_{dead} = 0.48 \pm 0.13$ nm, as shown in **Figure S1**. Note that the $\frac{M}{A}$ of Pt$_{0.69}$Cr$_{0.31}$(5)/Co($t_{Co}$)/MgO(2) with $t_{Co}$=1.2 nm shows good agreement with the trend of Pt/Co/MgO, which indicates that Cr dopants cause a negligible degradation on $M_s$ of Co layer in the Cr-doped Pt/Co/MgO heterostructures. Moreover, a small difference of $\frac{M}{A}$ is observed in Pt$_x$Cr$_{1-x}$(5)/Co(1.2)/MgO(2): $\frac{M}{A} = 6.26 \times 10^{-5}$ emu/cm$^2$ for pure Pt; $\frac{M}{A} = 8.30 \times 10^{-5}$ emu/cm$^2$ for Pt$_{0.69}$Cr$_{0.31}$. This difference is contributed from the weak ferromagnetism of Pt-Cr underlayer, confirmed by the VSM hysteresis loop measurement on a Pt$_{0.69}$Cr$_{0.31}$(5) sample: $\frac{M}{A} = 2.04 \times 10^{-5}$ emu/cm$^2$, giving $M_s = 40.88$ emu/cm$^3$ for Pt$_{0.69}$Cr$_{0.31}$.

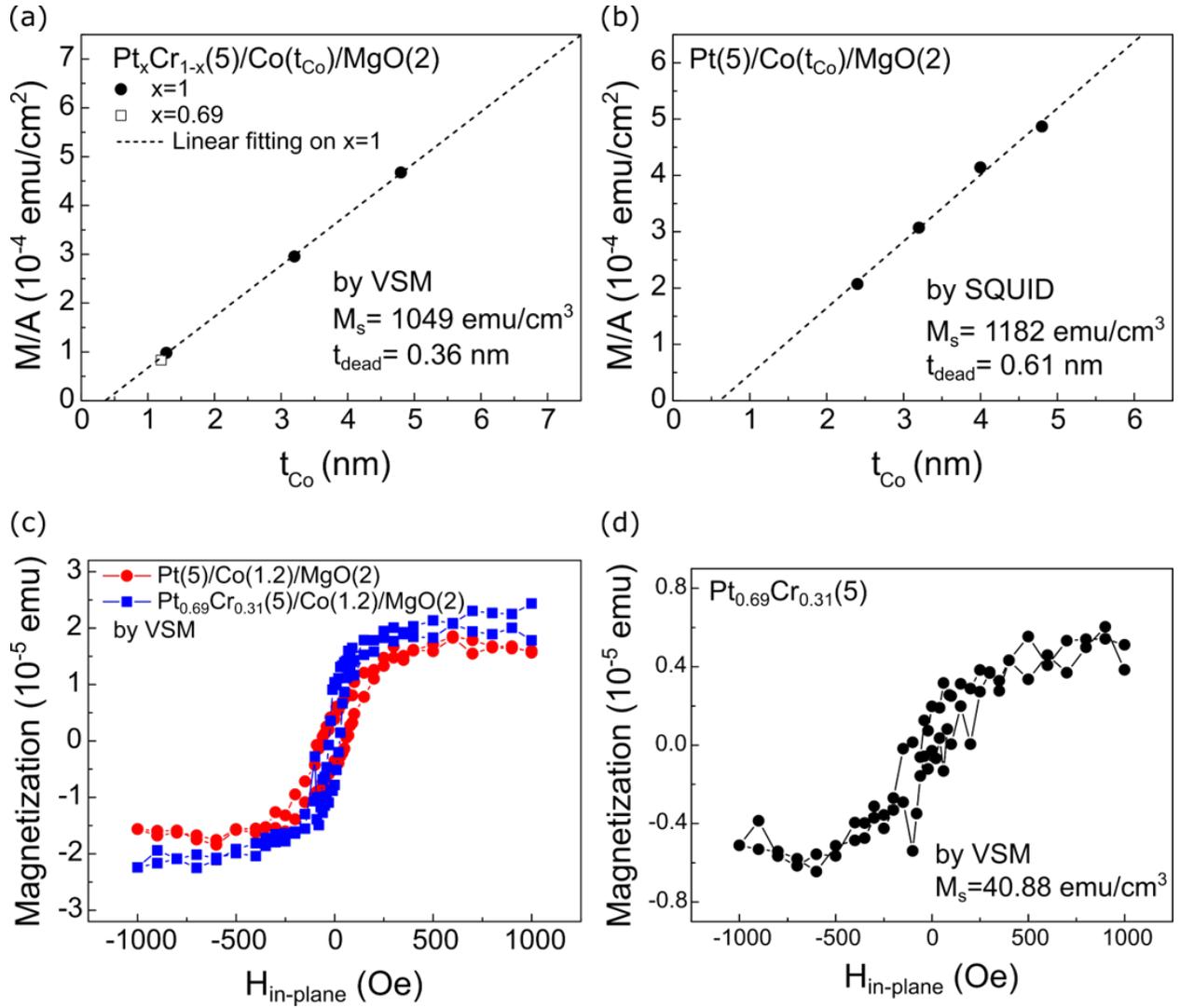

**Figure S1.** Magnetic properties of Pt-Cr/Co/MgO heterostructures. (a, b) $\frac{M}{A}$ versus $t_{Co}$ for Pt/Co/MgO characterized via VSM and SQUID. (c, d) Measured magnetization of Pt (5)/Co(1.2)/MgO(2), Pt$_{0.69}$Cr$_{0.31}$ (5)/Co(1.2)/MgO(2) and Pt$_{0.69}$Cr$_{0.31}$(5) by VSM.

### 3.2. Pt/CoFeB/MgO

The $\frac{M}{A}$ of Pt(3)/CoFeB(2.3)/MgO(2) capped by Ta(2) are 1.89×10$^{-4}$ emu/cm$^2$ by VSM and 2.07×10$^{-4}$ emu/cm$^2$ by SQUID, resulting in an average $M_s = 860 \pm 55$ emu/cm$^3$ with a negligible $t_{\text{dead}}$. Moreover, from the SQUID results, the magnitude of effective anisotropy field ($H_k^{\text{eff}}$) is found to be ~ 8000 Oe, as shown in **Figure S2**. This $H_k^{\text{eff}}$ value will be used in the protocol of determination of current-induced effective fields from the planar Hall effect curve shift measurement, which will be discussed in details in the following sections.

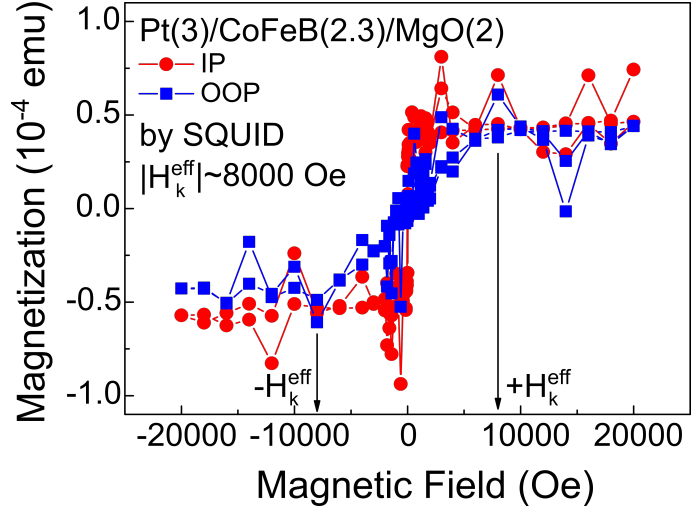

**Figure S2.** The in-plane (IP) and out-of-plane (OOP) hysteresis loop of a Pt(3)/CoFeB(2.3)/MgO(2) structure as characterized via SQUID, where $H_k^{\text{eff}}$ is the effective anisotropy field.

## 4. Planar Hall effect curve shift measurement

### 4.1. Magnetization direction deviation due to the current-induced fields

This part of formula derivation is modified from the work of S. J. Yun *et al.*[S1] Based on Stoner–Wohlfarth model, the total magnetic static energy ($E_{\text{total}}$) including anisotropy energy and Zeeman energy can be expressed as

$$E_{\text{total}} = -K^{\text{eff}} \cos^2 \theta_M - M_s \hat{m} \cdot \overrightarrow{H_{\text{ext}}}, \tag{S5}$$

where $K^{\text{eff}} = K_1 - N_d \frac{M_s^2}{2}$ is the effective anisotropy energy, including the first-order anisotropy energy and the demagnetization energy[S2]; $\hat{m} = (\sin \theta_M \cos \varphi_M, \sin \theta_M \sin \varphi_M, \cos \theta_M)$ is the magnetization unit vector; $\theta_M$ and $\varphi_M$ respectively are the polar and azimuthal angles of $\hat{m}$; $M_s$ is the saturation magnetization; $\overrightarrow{H_{\text{ext}}} = (H_x, H_y, H_z)$ is the applied external field vector. The normalized magnetic static energy ($\varepsilon_{\text{total}} \equiv \frac{E_{\text{total}}}{M_s}$) is modified as the following equation:

$$\varepsilon_{\text{total}} = -\frac{K^{\text{eff}}}{M_s} \cos^2 \theta_M - \hat{m} \cdot \overrightarrow{H_{\text{ext}}}. \tag{S6}$$

Under the equilibrium condition,

$$\frac{\partial \varepsilon_{\text{total}}}{\partial \theta_M} = 0 = H_k^{\text{eff}} \cos \theta_M \sin \theta_M + (-H_x \cos \theta_M \cos \varphi_M - H_y \cos \theta_M \sin \varphi_M + H_z \sin \theta_M),$$

$$\frac{\partial \varepsilon_{\text{total}}}{\partial \varphi_M} = 0 = (H_x \sin \varphi_M - H_y \cos \varphi_M) \sin \theta_M, \tag{S7}$$

where $H_k^{\text{eff}} \equiv \frac{2K^{\text{eff}}}{M_s}$ is the effective anisotropy field. Then, to estimate the angle deviation due to the current-induced field ($\overrightarrow{\Delta H} = (\Delta H_x, \Delta H_y, \Delta H_z)$), the derivatives of **Equation S7** with respect to $H_{i=x,y,z}$ should be zero,

$$\frac{\partial}{\partial H_i}\left(\frac{\partial \varepsilon_{\text{total}}}{\partial \theta_M}\right) = 0$$

$$= \left[H_k^{\text{eff}} \cos 2\theta_M \right.$$

$$\left. + \left(H_x \sin\theta_M \cos\varphi_M + H_y \sin\theta_M \sin\varphi_M + H_z \cos\theta_M\right)\right]\frac{\partial \theta_M}{\partial H_i}$$

$$+ \left(H_x \cos\theta_M \sin\varphi_M - H_y \cos\theta_M \cos\varphi_M\right)\frac{\partial \varphi_M}{\partial H_i} - f_i,$$

$$\frac{\partial}{\partial H_i}\left(\frac{\partial \varepsilon_{\text{total}}}{\partial \varphi_M}\right) = 0$$

$$= \left(H_x \sin\varphi_M - H_y \cos\varphi_M\right)\cos\theta_M \frac{\partial \theta_M}{\partial H_i}$$

$$+ \left(H_x \cos\varphi_M + H_y \sin\varphi_M\right)\sin\theta_M \frac{\partial \varphi_M}{\partial H_i} - g_i,$$

$$f_i \equiv (\cos\theta_M \cos\varphi_M, \cos\theta_M \sin\varphi_M, -\sin\theta_M),$$

$$g_i \equiv (-\sin\theta_M \sin\varphi_M, \sin\theta_M \cos\varphi_M, 0). \tag{S8}$$

Now if the in-plane anisotropy is negligible (causing $\varphi_M = \varphi_H$), combining with $\overrightarrow{H_{\text{ext}}} = (H_x, H_y, H_z) = H_{\text{ext}}(\sin\theta_H \cos\varphi_H, \sin\theta_H \sin\varphi_H, \cos\theta_H)$, the above equation can be modified as

$$\frac{\partial}{\partial H_i}\left(\frac{\partial \varepsilon_{\text{total}}}{\partial \theta_{\text{M}}}\right) = 0$$

$$= \left[H_k^{\text{eff}} \cos 2\theta_{\text{M}}\right.$$

$$+ H_{\text{ext}}(\sin\theta_{\text{H}}\cos\varphi_{\text{H}}\sin\theta_{\text{M}}\cos\varphi_{\text{H}} + \sin\theta_{\text{H}}\sin\varphi_{\text{H}}\sin\theta_{\text{M}}\sin\varphi_{\text{H}}$$

$$\left.+ \cos\theta_{\text{H}}\cos\theta_{\text{M}})\right]\frac{\partial \theta_{\text{M}}}{\partial H_i}$$

$$+ H_{\text{ext}}(\sin\theta_{\text{H}}\cos\varphi_{\text{H}}\cos\theta_{\text{M}}\sin\varphi_{\text{H}} - \sin\theta_{\text{H}}\sin\varphi_{\text{H}}\cos\theta_{\text{M}}\cos\varphi_{\text{H}})\frac{\partial \varphi_{\text{M}}}{\partial H_i}$$

$$- f_i = \left[H_k^{\text{eff}} \cos 2\theta_{\text{M}} + H_{\text{ext}}(\sin\theta_{\text{H}}\sin\theta_{\text{M}} + \cos\theta_{\text{H}}\cos\theta_{\text{M}})\right]\frac{\partial \theta_{\text{M}}}{\partial H_i} - f_i$$

$$= \left[H_k^{\text{eff}} \cos 2\theta_{\text{M}} + H_{\text{ext}}\cos(\theta_{\text{M}} - \theta_{\text{H}})\right]\frac{\partial \theta_{\text{M}}}{\partial H_i} - f_i,$$

$$\frac{\partial}{\partial H_i}\left(\frac{\partial \varepsilon_{\text{total}}}{\partial \varphi_{\text{M}}}\right) = 0$$

$$= H_{\text{ext}}(\sin\theta_{\text{H}}\cos\varphi_{\text{H}}\sin\varphi_{\text{H}} - \sin\theta_{\text{H}}\sin\varphi_{\text{H}}\cos\varphi_{\text{H}})\cos\theta_{\text{M}}\frac{\partial \theta_{\text{M}}}{\partial H_i}$$

$$+ H_{\text{ext}}(\sin\theta_{\text{H}}\cos\varphi_{\text{H}}\cos\varphi_{\text{H}} + \sin\theta_{\text{H}}\sin\varphi_{\text{H}}\sin\varphi_{\text{H}})\sin\theta_{\text{M}}\frac{\partial \varphi_{\text{M}}}{\partial H_i} - g_i$$

$$= H_{\text{ext}}\sin\theta_{\text{H}}\sin\theta_{\text{M}}\frac{\partial \varphi_{\text{M}}}{\partial H_i} - g_i,$$

$$f_i \equiv (\cos\theta_{\text{M}}\cos\varphi_{\text{H}}, \cos\theta_{\text{M}}\sin\varphi_{\text{H}}, -\sin\theta_{\text{M}}),$$

$$g_i \equiv (-\sin\theta_{\text{M}}\sin\varphi_{\text{H}}, \sin\theta_{\text{M}}\cos\varphi_{\text{H}}, 0), \tag{S9}$$

giving the total angle deviation due to $\overrightarrow{\Delta H}$ expressed as

$$\Delta\theta_{\text{M}} = \sum_{i=x,y,z}\frac{\partial \theta_{\text{M}}}{\partial H_i}\Delta H_i = \frac{(\Delta H_x \cos\varphi_{\text{H}} + \Delta H_y \sin\varphi_{\text{H}})\cos\theta_{\text{M}} - \Delta H_z \sin\theta_{\text{M}}}{H_k^{\text{eff}}\cos 2\theta_{\text{M}} + H_{\text{ext}}\cos(\theta_{\text{M}} - \theta_{\text{H}})},$$

$$\Delta\varphi_{M} = \sum_{i=x,y,z} \frac{\partial \varphi_M}{\partial H_i} \Delta H_i = \frac{\left(-\Delta H_x \sin\varphi_H + \Delta H_y \cos\varphi_H\right)\sin\theta_M}{H_{ext}\sin\theta_H \sin\theta_M}$$

$$= \frac{-\Delta H_x \sin\varphi_H + \Delta H_y \cos\varphi_H}{H_{ext}\sin\theta_H}. \tag{S10}$$

### 4.2. Transverse resistance changing due to current-induced magnetization deviation

The transverse resistance ($R_{xy}$) which includes the contributions from anomalous Hall effect (AHE), planar Hall effect (PHE) and anomalous Nernst effect (ANE) can be expressed as

$$R_{xy} = R_{AHE}m_z + 2R_{PHE}m_x m_y + I\alpha_{ANE}\nabla T m_x$$

$$= R_{AHE}\cos\theta_M + 2R_{PHE}\sin^2\theta_M \cos\varphi_M \sin\varphi_M + I\alpha\nabla T \sin\theta_M \cos\varphi_M, \tag{S11}$$

where $R_{AHE(PHE)}$ is half of the maximum transverse resistance change due to AHE or PHE; $I$ is the applied current along $x$-direction; $\alpha_{ANE}$ is the ANE coefficient; $\nabla T$ is the thermal gradient along positive $z$-direction. For an in-plane magnetized device with an injected $I_{DC}$ along $x$-direction, the magnetization would deviate from its original equilibrium direction ($\theta_M^0$, $\varphi_M^0$) due to the presence of $\overrightarrow{\Delta H}$:

$$\theta_M = \theta_M^0 + \Delta\theta_M,$$

$$\varphi_M = \varphi_M^0 + \Delta\varphi_M, \tag{S12}$$

where $\theta_M$ is the exact polar angle of $\hat{m}$; $\theta_M^0$ is the equilibrium polar angle under no influence from $\overrightarrow{\Delta H}$; $\Delta\theta_M$ is the polar angle deviation due to $\overrightarrow{\Delta H}$. The notation is the same for azimuthal angle. Therefore, the $R_{xy}$ under magnetization deviation can be estimated by a first-order approximation:

$$R_{xy} = R_{AHE}\cos\theta_M + 2R_{PHE}\sin^2\theta_M \cos\varphi_M \sin\varphi_M +$$

$$I\alpha\nabla T \sin\theta_M \cos\varphi_M \sim R_{AHE}(\cos\theta_M^0 - \Delta\theta_M \sin\theta_M^0) + 2R_{PHE}[\sin^2\theta_M^0 \cos\varphi_M^0 \sin\varphi_M^0 +$$

$$\sin\theta_M^0 (\Delta\theta_M \cos\theta_M^0 \sin 2\varphi_M^0 + \Delta\varphi_M \sin\theta_M^0 \cos 2\varphi_M^0)] + I\alpha\nabla T \sin\theta_M^0 \cos\varphi_M^0. \quad (S13)$$

Now, consider the scenario where an IMA device with negligible in-plane anisotropy, such that $\theta_M^0 = 90°$ and $\varphi_M^0 = \varphi_H$:

$$R_{xy} \sim R_{AHE}(-\Delta\theta_M) + 2R_{PHE}\left(\frac{1}{2}\sin 2\varphi_H + \Delta\varphi_M \cos 2\varphi_H\right) + I\alpha\nabla T \cos\varphi_M^0$$

$$= R_{AHE}(-\Delta\theta_M) + R_{PHE}(\sin 2\varphi_H + 2\Delta\varphi_M \cos 2\varphi_H) + I\alpha\nabla T \cos\varphi_M^0. \quad (S14)$$

Based on the SHE scenario (which includes the current-induced effective fields from the damping-like and field-like spin-orbit torques) combined with current-induced Oersted field, the overall current-induced field can be expressed as:

$$\overrightarrow{\Delta H} = (\Delta H_x, \Delta H_y, \Delta H_z) = (-\Delta H_{DL}m_z, -\Delta H_{FL} - \Delta H_{Oe}, \Delta H_{DL}m_x) =$$

$$(0, -\Delta H_{FL} - \Delta H_{Oe}, \Delta H_{DL}\cos\varphi_H), \quad (S15)$$

where $\Delta H_{i=x,y,z}$ are the component of $\overrightarrow{\Delta H}$ along $i$-direction; $\Delta H_{\text{DL(FL)}}$ is the amount of current-induced damping(field)-like spin-orbit field; $\Delta H_{\text{Oe}}$ is the Oersted field. Combing **Equation S10** and **Equation S15**, current-induced magnetization deviations can be expressed as

$$\Delta\theta_M = \frac{-\Delta H_z}{-H_k^{\text{eff}} + H_{\text{ext}}} = \frac{-\Delta H_{\text{DL}}}{-H_k^{\text{eff}} + H_{\text{ext}}} \cos\varphi_H,$$

$$\Delta\varphi_M = \frac{-\Delta H_{\text{FL}} - \Delta H_{\text{Oe}}}{H_{\text{ext}}} \cos\varphi_H. \tag{S16}$$

Therefore, the angle-dependent transverse resistance is

$$\begin{aligned}
R_{xy} &\sim \left[R_{\text{AHE}}\left(\frac{\Delta H_{\text{DL}}}{-H_k^{\text{eff}} + H_{\text{ext}}}\right) + I\alpha\nabla T\right]\cos\varphi_H \\
&+ R_{\text{PHE}}\left(\sin 2\varphi_H + \frac{-2(\Delta H_{\text{FL}} + \Delta H_{\text{Oe}})}{H_{\text{ext}}}\cos\varphi_H \cos 2\varphi_H\right) \\
&= \left[R_{\text{AHE}}\left(\frac{\Delta H_{\text{DL}}}{-H_k^{\text{eff}} + H_{\text{ext}}}\right) + I\alpha\nabla T\right]\cos\varphi_H \\
&+ R_{\text{PHE}}\left\{\sin 2\varphi_H + \left[\frac{-2(\Delta H_{\text{FL}} + \Delta H_{\text{Oe}})}{H_{\text{ext}}}\right](2\cos^3\varphi_H - \cos\varphi_H)\right\} \\
&= R_{\text{PHE}}\sin 2\varphi_H + \left[R_{\text{AHE}}\left(\frac{\Delta H_{\text{DL}}}{-H_k^{\text{eff}} + H_{\text{ext}}}\right) + I\alpha\nabla T\right]\cos\varphi_H \\
&+ R_{\text{PHE}}\left[\frac{-2(\Delta H_{\text{FL}} + \Delta H_{\text{Oe}})}{H_{\text{ext}}}\right](2\cos^3\varphi_H - \cos\varphi_H),
\end{aligned} \tag{S17}$$

which contains a current-independent component, $\sin 2\varphi_H$, and two current-dependent components, $\cos\varphi_H$ and $(2\cos^3\varphi_H - \cos\varphi_H)$.

In fact, the thermal term $I\alpha\nabla T \sin\theta_M \cos\varphi_M \sim I\alpha\nabla T(\sin\theta_M^0 \cos\varphi_M^0 - \Delta\varphi_M \sin\theta_M^0 \sin\varphi_M^0)$, which includes the contribution from $\Delta H_{FL}$ and $\Delta H_{Oe}$. Nevertheless, if the materials system possesses small FL-SOT efficiency and/or the experimental setup generates small $\Delta H_{Oe}$ (either using resistive spin current source or applying relatively small current amount), $I\alpha\nabla T \sin\theta_M \cos\varphi_M \sim I\alpha\nabla T \sin\theta_M^0 \cos\varphi_M^0$ would be an acceptable approximation.

**4.3. Planar Hall effect (PHE) curve-shift measurement**

First, we define the average and the difference of $R_{xy}$ under opposite current directions as follows,

$$\overline{R_{xy}} \equiv \frac{1}{2}[R_{xy}(+I) + R_{xy}(-I)],$$

$$\Delta R_{xy} \equiv \frac{1}{2}[R_{xy}(+I) - R_{xy}(-I)], \quad (S18)$$

where $R_{xy}(\pm I)$ is the measured $R_{xy}$ under $\pm I_{DC}$. Considering the nature that $R_{PHE}$ is independent of the applied current direction whereas $\Delta H_{DL,FL,Oe}$ and $I\alpha\nabla T$ depend on the current direction, therefore,

$$\overline{R_{xy}} = R_{PHE} \sin 2\varphi_H,$$

S-14

$$\Delta R_{xy} = \left[ R_{\text{AHE}} \left( \frac{\Delta H_{\text{DL}}}{-H_k^{\text{eff}} + H_{\text{ext}}} \right) + I\alpha \nabla T \right] \cos \varphi_H$$

$$+ R_{\text{PHE}} \left[ \frac{-2(\Delta H_{\text{FL}} + \Delta H_{\text{Oe}})}{H_{\text{ext}}} \right] (2\cos^3 \varphi_H - \cos \varphi_H)$$

$$= R^{\text{DL+AN}} \cos \varphi_H + R^{\text{FL+Oe}} (2\cos^3 \varphi_H - \cos \varphi_H). \tag{S19}$$

$\Delta H_{\text{DL}}$ can be extracted from the slope of the linear fitting of $R^{\text{DL+A}}$ versus $\frac{1}{-H_k^{\text{eff}} + H_{\text{ext}}}$ data, where $R_{\text{AHE}}$ at $\theta_H \sim 90°$ is estimated from $R_{\text{AHE}} \sim -\frac{\partial \overline{R_{xy}}}{\partial \Delta \theta_H} = -\frac{\partial \overline{R_{xy}}}{\partial \theta_H} \frac{\partial \theta_H}{\partial \Delta \theta_H} = -\frac{\partial \overline{R_{xy}}}{\partial \theta_H} \left( \frac{-H_k^{\text{eff}} + H_{\text{ext}}}{H_{\text{ext}}} \right)$ by applying an external field in the x-z plane ($\varphi_H = 0°$) with varying its polar angle from $\theta_H = 85°$ to $95°$. Take the measurement on $Pt_{0.69}Cr_{0.31}(3)/CoFeB(2.3)/MgO(2)$ for example, $-\frac{\partial \overline{R_{xy}}}{\partial \theta_H} = 0.00632$ $\Omega/° = 0.362$ $\Omega/\text{rad}$ under $H_{\text{ext}} = 1000$ Oe and $|I_{\text{DC}}| = 3$ mA, giving $R_{\text{AHE}} \sim 3.26$ $\Omega$, as shown in **Figure S3**.

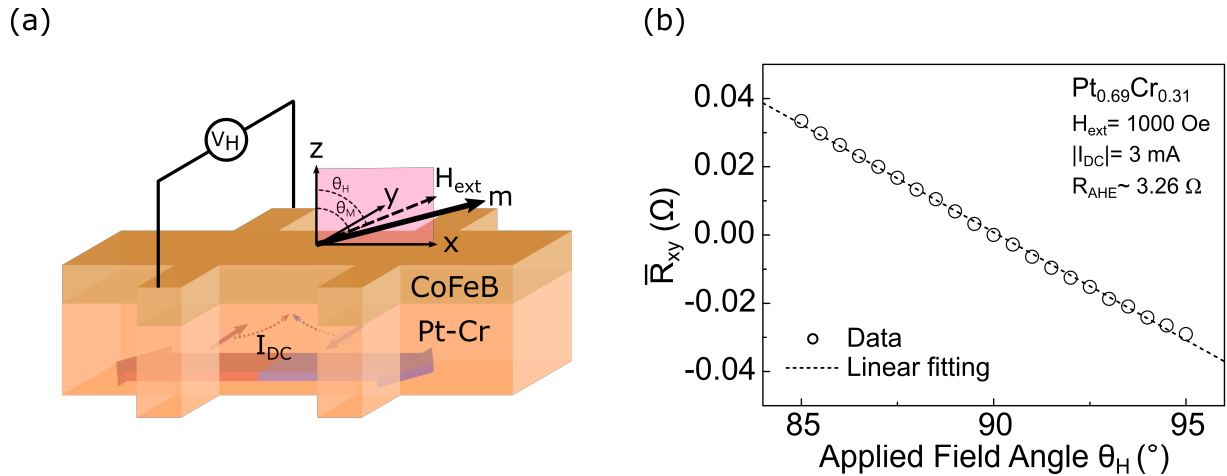

**Figure S3.** The details of determining $R_{\text{AHE}}$ in PHE curve-shift measurement. (a) Schematics of experimental setup. (b) Representative data of $\overline{R_{xy}}$ under scanning $H_{\text{ext}} = 1000$ Oe in x-z plane.

Also, $(\Delta H_{FL} + \Delta H_{Oe})$ can be extracted from the term $\frac{-R^{FL+Oe}}{2\overline{R_{xy}}} H_{ext}$. In the case of Pr$_{0.69}$Cr$_{0.31}$, under $H_{ext} = 1000$ Oe, $|R^{FL+Oe}| \sim 10$ mΩ and $|\overline{R_{xy}}| \sim 1$ Ω, giving a negligible $\frac{-\Delta H_{FL} - \Delta H_{Oe}}{H_{ext}}$. Therefore, the assumption on $I\alpha\nabla T \sin\theta_M \cos\varphi_M \sim I\alpha\nabla T \sin\theta_M^0 \cos\varphi_M^0$ is still reasonable.

## 5. Power consumption estimation of a prototypic SOT-MRAM device

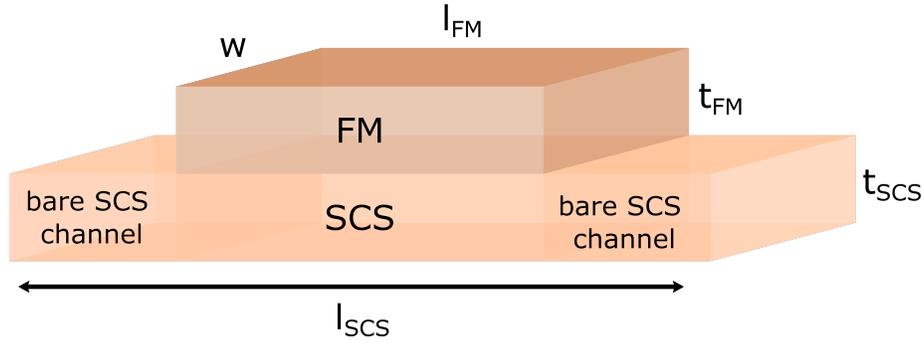

**Figure S4.** Schematic and parameter definition of a prototypic SOT-MRAM

A fair comparison on the power consumption of SOT switching among different spin current sources (SCSs) can be obtained based on the information of the SCS DL-SOT efficiencies ($\xi_{DL}$) and resistivities ($\rho_{SCS}$) with the assumption of employing the same device geometry and ferromagnetic materials (FM) as the memory storage layer. For simplicity we choose a simple SCS( $t_{SCS} = 5$ nm )/FM( $t_{FM} = 1$ nm ) bilayer structure with $w = 2$ μm, $l_{FM} = 2$ μm, $l_{SCS\ channel} = 3$ μm. The schematics of this prototypic device is shown in Figure S4. For a perpendicularly-magnetized SOT-MRAM under the presence of domain wall, the critical switching current density ($J_c$) and the power consumption ($P$) for writing a single bit are

$$J_\text{c} = \frac{C}{\xi_\text{DL}},$$

$$P = (1+s)^2 C^2 \rho_\text{SCS} \xi_\text{DL}^{-2} V_\text{bare channel} + (1+s) C^2 \rho_\text{SCS} \xi_\text{DL}^{-2} V_\text{SCS}, \tag{S20}$$

where $C = \left(\frac{2}{\pi}\right)\left(\frac{2e}{\hbar}\right)\mu_0 M_\text{s} t_\text{FM} H_\text{c}$, $\mu_0$ is the vacuum permittivity; $M_\text{s}$ is the saturation magnetization; $H_\text{c}$ is the coercivity; $V_\text{bare channel} \equiv w t_\text{SCS}(l_\text{SCS channel} - l_\text{FM})$ is the bare SCS channel volume; $V_\text{SCS} \equiv w t_\text{SCS} l_\text{FM}$ is the SCS volume underneath the FM layer.